\documentclass[
bibliography=totoc,    
english]
{scrartcl}           

\usepackage[T1]{fontenc}           
\usepackage[utf8]{inputenc}      
\usepackage[english]{babel}            
\usepackage{csquotes}            

\usepackage{lmodern}              
\usepackage{microtype}         

\usepackage{
	amsfonts,  
	amsmath,
	amssymb, 
	amsthm,
	array,
	authblk,
	bbding,
	bbm,
	booktabs,
	calc,
	caption,
	color, 
	enumitem,
	float,
	graphicx,   
	here,
	lipsum,
	listings,
	longtable,  
	mathtools,
	multirow,
	pdfpages,
	pifont,
	placeins,
	ragged2e,
	rotating,
	scrlayer-scrpage,
	setspace,
	subcaption,                           
	tabularx,
}

\usepackage{
	textcomp,
	todonotes
	}

\usepackage[
	style			= apa,
	citestyle		= authoryear,
	firstinits,
	hyperref		= true, 
	natbib 			= true,
	backref			= false, 
	citetracker		= true, 
	maxcitenames	= 2,
	maxbibnames		= 99,
	doi				= true, 
	isbn			= false, 
	useprefix		= false,
	dashed 			= false,
	uniquename 		= false,
	backend=biber]{biblatex}
	
\bibliography{myBib_sae.bib,bib_R_packages.bib}

\DeclareNameAlias{sortname}{last-first} 

\usepackage{geometry}
\pdfoutput=1

\begin{document}                   
	
	\date{\today}
	
	\title{Evaluation of small-area estimation methods for mortality schedules}
	
	\author[1,2]{Esther Denecke\thanks{Corresponding author: denecke@demogr.mpg.de}}
	\author[3]{Pavel Grigoriev}
	\author[2,4]{Roland Rau}
	\affil[1]{\small{Department of Digital and Computational Demography, Max Planck Institute for Demographic Research, Rostock, Germany}}
	\affil[2]{\small{Department of Sociology \& Demography, University of Rostock, Germany}}
	\affil[3]{\small{Federal Institute for Population Research (BiB),  Wiesbaden, Germany}}	
	\affil[4]{\small{Laboratory of Statistical Demography, Max Planck Institute for Demographic Research, Rostock, Germany}}

	\maketitle                       

	\vspace*{-0.5cm}
	
	\section*{Abstract}
	Mortality patterns at a subnational level or across subpopulations are often used to examine the health of a population. In small populations, however, death counts are erratic. To deal with this problem, demographers have proposed different methods but it is unclear how these methods relate to each other. We aim to provide guidance. First, we review recent demographic small-area methods for mortality schedules. Second, we evaluate three methodological approaches using simulated data. We show that there is considerable variability in the performance across ages and subpopulations/regions and that this performance can depend on the choice of incorporated demographic knowledge. \\
	\textbf{Keywords}: comparative assessment, demographic methods, mortality, simulation study, small-area estimation

	\section{Introduction}
	
	Assessing mortality in different regions or subpopulations is important for understanding population health and for designing target health policies. Misguided health policies are problematic for at least two reasons: They are expensive and do not help those in need. While the estimation of mortality rates is fairly straightforward for large populations, problems arise with small populations as the underlying death counts are subject to high variability. In order to deal with this variability, data are often pooled across space or time leading to a loss of information. Alternatively, researchers have explored both traditional life table methods as well as different types of models to estimate mortality rates and life expectancy in small populations \parencites{Congdon2009ISR, Congdon2004JoAS, Congdon2013JoGS, EayresWilliams2004JoECH, JonkerEtAl2012AJoE, RashidEtAl2021TLPH, RichardsonEtAl2009SSM, ScherbovEdiev2011DR, StephensEtAl2013BO, TosonBaker2003, VoulgarakiEtAl2014SIM}. Recent developments also include \textit{demographic small-area estimation methods} \parencites{AlexanderEtAl2017D, GonzagaSchmertmann2016RBdEdP, RauSchmertmann2020DAO, SchmertmannGonzaga2018D, Schmertmann2021DR}. These \textit{demographic methods} make---in our understanding---use of demographic properties such as the age-specific shape of mortality. 
	
	Making use of knowledge about the shape of human mortality curves has a long tradition in demography. The Brass relational model \parencite{Brass1971BAD} relates the (logit) survivorship rates of a target population to the (logit) survivorship rates of a so-called \textit{standard} population by means of two parameters. In a different fashion, \textcite{LeeCarter1992JotASA} model and project age-specific mortality rates by using the first right and left singular values obtained by decomposing a matrix of centered log mortality rates. Demographic small-area estimation methods follow these traditions by using knowledge about the general pattern of age-specific mortality.
	
	While some of these recent demographic methods have been compared to other methods by means of simulation studies, it is generally unclear how the methods relate to each other and how to choose one in practice. For example, \textcite{GonzagaSchmertmann2016RBdEdP} compare frequentist TOPALS regression to indirect standardization whereas \textcite{Schmertmann2021DR} compares Demographic Splines (D-splines) to penalized splines (P-splines). \textcite{GonzagaSchmertmann2016RBdEdP} also demonstrate that frequentist TOPALS regression is insensitive to the choice of incorporated demographic knowledge (=standard). Each comparison comes with different set-ups in terms of input data, underlying mortality rates, population sizes, number of repetitions, and performance metrics limiting the comparability across studies. Additionally, previous studies often average the results over several simulated regions/populations or years potentially masking underlying variability. Lastly, as complex methods might not be used in practice \parencites{SevcikovaRaftery2021JoOS, Wilson2018G}, it is important to not only assess the performance but to also look at data requirements and ease of use. Therefore, we aim (i) to review recent demographic small-area estimation methods for mortality schedules and (ii) to empirically evaluate three of them on simulated data.
	
	For the review of methods, we not only describe the methods but also present a qualitative comparison in terms of data requirements and ease of use. Our simulation studies allow us to study the interplay of exposure size (=person-years lived) and incorporated demographic knowledge across different regions/subpopulations for estimating age-specific mortality rates and life expectancy. The simulation study itself contributes to the literature in several ways. 
	First, we evaluate frequentist TOPALS regression \parencite{GonzagaSchmertmann2016RBdEdP}, D-splines \parencite{Schmertmann2021DR}, and a Bayesian hierarchical model using right singular vectors as covariates \parencites{AlexanderEtAl2017D} within a systematic and (as far as possible) unified framework.
	Second, we study the ability to estimate age-specific log mortality rates with respect to five different performance measures, namely bias, empirical standard error, root mean squared error as well as coverage and width of the uncertainty intervals as often the interplay of these measures is of importance \parencite{MorrisEtAl2019SiM}. 
	Third, we provide detailed results as we assess the performance for age-specific mortality rates instead of averaging over all ages and/or areas/subpopulations. We also provide distinct results for life expectancy by separating out different regions instead of averaging over them. 
	Fourth, we evaluate frequentist TOPALS regression and D-splines for both single year as well as age-grouped input data. 
	
	The aim of our study is explicitly not to find \textit{the best method} as there usually is no universally best or worst method \parencite{StroblLeisch2022BJ}. In fact, \textcite[2]{StroblLeisch2022BJ} go as far as saying that the search for a best method is \enquote{an ill-posed question}. Instead, we aim to provide a state-of-the art overview summarizing the advantages and disadvantages of different demographic small-area estimation methods. We hope that this will help both the applied researcher in choosing a method as well as the methodologist by pointing out strengths and weaknesses of current methods.
	
	\section{Review of Methods} 
	\label{sec: methods}
	
	There are three broad approaches of demographic small area-estimation methods for full mortality schedules. These approaches are based on how knowledge about human mortality curves (demographic knowledge) is incorporated. 
	In the first approach, one standard schedule, i.e.~a (smoothed) log mortality schedule that is representative of the shape of human mortality is integrated into the model. This approach is composed of (i) TOPALS regression models \parencites{GonzagaSchmertmann2016RBdEdP, SchmertmannGonzaga2018D, RauSchmertmann2020DAO} in which deviations from a standard schedule are modeled by means of (linear) splines and (ii) Bayesian Dynamic models \parencite{OliveiraEtAl2021} where the standard is used as a covariate. 
	The second approach is comprised of D-splines (\enquote{Demographic Splines}) \parencite{Schmertmann2021DR}. In D-splines, the penalty term of P-splines \parencite{EilersMarx1996StS} is altered to make use of demographic knowledge extracted from a database of log mortality rates.
	Models following the third approach use right-singular values obtained by decomposing a matrix of log mortality rates as covariates within a Bayesian hierarchical model (thereafter called \textit{SVD-model}) \parencite{AlexanderEtAl2017D}. \textcite{DharamshiEtAl2021PAA} propose an extension to \textcite{AlexanderEtAl2017D} modeling mortality schedules of multiple subgroups at once. 
	Each of these approaches can be composed of both frequentist and Bayesian methods.
	
	In the following, we briefly describe the main ideas of (i) TOPALS regression, (ii) D-splines, and (iii) SVD-models, i.e.~one type of model from each approach. We conclude this section with an overview of the methods with respect to data requirements as well as their outputs. We use $x = 0, \dots, \left( \omega - 1 \right)$ to denote single year ages such that $ \left( \omega - 1 \right) $ is the oldest age. Similarly, we use $g = 1, \dots, G$ for age groups. Age groups are in $5$-year intervals (except for the youngest ages), i.e. $<1, 1-4, 5-9, \dots, 95-99$. Thus, $g = 1$ refers to the youngest age group (age $<1$). All of the methods described assume that age-specific deaths $D_x$ are Poisson distributed with exposures $N_x$ and death rates $m_x$, $D_x \sim \text{Poisson} \left(m_x N_x\right)$. Consequently, also age-grouped deaths are Poisson distributed, $D_g \sim \text{Poisson} \left(m_g N_g\right)$. Finally, $\mathbf{1}_{\omega}$ denotes the $\omega$-dimensional column vector of ones.
	
	\section*{\normalfont \textit{TOPALS regression}}
	
	TOPALS (\enquote{tool for projecting age patterns using linear splines}) is a relational model first introduced to estimate and project age-specific fertility rates \parencite{Beer2011DR} and probabilities of dying \parencite{Beer2012DR}. \textcite{GonzagaSchmertmann2016RBdEdP} embed TOPALS within a frequentist Poisson regression framework (=frequentist TOPALS regression) where deviations from a representative standard log mortality schedule are modeled by means of (linear) splines. \textcite{SchmertmannGonzaga2018D} transfer this approach into a Bayesian framework that can be used with data that is incomplete in terms of death registration. Lastly, \textcite{RauSchmertmann2020DAO} use a Bayesian TOPALS approach to estimate age-specific mortality rates in each German district. They incorporate a hierarchical structure into the model accounting for the fact that -- roughly speaking -- districts are nested within states which in turn are located within Germany\footnote{This is a rough description only as some German states have an additional administrative layer (\textit{Regierungsbezirke}). For a detailed description including the hierarchical structure see the supplementary material of \textcite{RauSchmertmann2020DAO}.}. Additionally, they incorporate knowledge on the differences in mortality schedules by women and men by means of a prior, the \enquote{sex-difference prior}. These three TOPALS models penalize large differences in mortality rates in neighboring ages, thus enforcing smoothness of the mortality schedules.
	
	There are two key differences between the TOPALS regression models described above. First, in the frequentist version \parencite{GonzagaSchmertmann2016RBdEdP} and the first Bayesian application \parencite{SchmertmannGonzaga2018D} mortality schedules are estimated for different areas independently of each other while the Bayesian model for Germany \parencite{RauSchmertmann2020DAO} borrows strength across space. Second, while all TOPALS regression models described above produce mortality rates for single year ages, the models differ in whether the input data, i.e.~the death counts and exposures, are given in single years \parencites{GonzagaSchmertmann2016RBdEdP, SchmertmannGonzaga2018D} or age groups \parencites{RauSchmertmann2020DAO}.
	
	In TOPALS, age-specific mortality rates are modeled as the sum of a standard schedule and (linear) B-Splines \parencite{GonzagaSchmertmann2016RBdEdP}
	$$
	\log \boldsymbol{m} = \boldsymbol{\lambda}_{\text{std}} + \boldsymbol{B} \boldsymbol{\alpha},
	$$
	where $\boldsymbol{m} = \left( m_0, \dots, m_{(\omega - 1)} \right)^T$ is a vector of age-specific mortality rates, $\boldsymbol{\lambda}_{\text{std}} \in \mathbb{R}^{\omega}$ a standard schedule which is typically a vector of \textit{representative} age-specific log mortality rates, $\boldsymbol{B} \in \mathbb{R}^{\omega \times S}$ is a B-spline basis matrix with $S$ linear B-splines and $\boldsymbol{\alpha} \in \mathbb{R}^{S}$ are the TOPALS parameters.
	
	In frequentist TOPALS regression, the TOPALS parameters $\boldsymbol{\alpha}$ are estimated via maximum likelihood \parencite{GonzagaSchmertmann2016RBdEdP}. While \textcite{GonzagaSchmertmann2016RBdEdP} show this for single year ages, \textcite{Schmertmann2022TopalsFittingWithGroupedData} illustrates the procedure and provides \texttt{R}-code for age-grouped input data. In the case of age-grouped input data, it is assumed that the age-grouped mortality rate $m_g$ is the average of the corresponding single year mortality rates $m_x$, i.e.~$ m_g = \frac{1}{|X_g|} \sum_{x \in X_g} m_x, $ where $X_g$ is the set of single year ages $x$ in age group $g$ and $|\cdot|$ denotes the cardinality of a set \parencites{RauSchmertmann2020DAO, Schmertmann2022TopalsFittingWithGroupedData}.
	
	In addition to producing age-specific point estimates $\log \hat{m}_x$, \textcite{GonzagaSchmertmann2016RBdEdP} outline the calculation of analytical standard errors for both the estimated log mortality rates, $\log \hat{m}_x$, and the TOPALS parameters $\hat{\boldsymbol{\alpha}}$. These analytical standard errors can be used to calculate pointwise confidence intervals.

	\section*{\normalfont \textit{D-splines}}
	
	D-splines (Demographic splines) are a type of P-splines (Penalized splines) that have been shown to have more favourable properties in small populations \parencite{Schmertmann2021DR}. Broadly, the idea is to construct residuals from a database of empirically observed mortality rates which are incorporated into the penalty term. Thus, the penalty term in D-splines contains information about the general form of human mortality schedules as opposed to P-splines which follow \enquote{generic arithmetical rules} \parencite[1089]{Schmertmann2021DR}. 
	
	In D-splines, a vector of age-specific log mortality rates is modeled as \parencite{Schmertmann2021DR}
	$$
	\log \boldsymbol{m} = \boldsymbol{S}\boldsymbol{\theta},
	$$
	where $\boldsymbol{m} = \left( m_0, \dots, m_{(\omega - 1)} \right)^T$ is a vector of age-specific mortality rates, $\boldsymbol{S}$ is a $\left(\omega \times K\right)$ B-spline basis matrix with $K$ cubic B-splines and $\boldsymbol{\theta} \in \mathbb{R}^{K}$ is a vector of spline coefficients.
	
	The spline coefficients $\boldsymbol{\theta}$ are estimated by maximizing the penalized log likelihood \parencite{Schmertmann2021DR}. The penalized log likelihood has the general form \parencite{Schmertmann2021DR}:
	$$
	Q\left(\boldsymbol{\theta}\right) = 
	\log L \left( \boldsymbol{\theta} \right) -  \frac{1}{2} \boldsymbol{\varepsilon}_{i}^{T} \left( \boldsymbol{\theta} \right) \left[ \hat{\boldsymbol{V}}_{i}^{-1} \right] \boldsymbol{\varepsilon}_{i} \left( \boldsymbol{\theta} \right),
	$$
	where the subtracted part is the penalty term with residual vectors $\boldsymbol{\varepsilon}_{i} \left( \boldsymbol{\theta} \right)$ and empirical covariance matrix $ \hat{\boldsymbol{V}}_{i}, i = 1, 2, 3$. The index $i$ refers to the type of estimator used (described below). The residual vector is defined as
	$$
	\boldsymbol{\varepsilon}_{i} \left( \boldsymbol{\theta} \right)
	= \boldsymbol{D}_{i}\boldsymbol{S}\boldsymbol{\theta} - \boldsymbol{c}_{i},
	$$
	where $\boldsymbol{S}\boldsymbol{\theta}$ are the age-specific log mortality rates and $\boldsymbol{D}_{i}$ and $\boldsymbol{c}_{i}$ are a precalibrated matrix and vector, respectively.
	
	The question that arises is what the predefined constants $\boldsymbol{D}_{i}$ and $\boldsymbol{c}_{i}$ as well as the empirical covariance matrix $\hat{\boldsymbol{V}}_{i}$ look like. \textcite{Schmertmann2021DR} proposes three different estimators: (i) The D-1 estimator which penalizes deviations from the slope of the empirically observed mortality schedules, (ii) the D-2 estimator which penalizes deviations from the curvature of the empirically observed mortality schedules, and (iii) the D-LC estimator which favours Lee-Carter fits given that they fit the observed data. The dimensions of $\boldsymbol{D}_{i}$ and $\boldsymbol{c}_{i}$ depend on the type of estimator used.\footnote{\textcite[1096]{Schmertmann2021DR} contains a typo regarding the dimension of $\boldsymbol{D}_{i}$. $\boldsymbol{D}_{i}$ (in notation of the paper matrix $A$) is declared as a square matrix which is incorrect as is correctly stated a page later \parencite[1097]{Schmertmann2021DR}.}
	
	In case of the D-1 estimator, the matrix $\boldsymbol{D}_{1} \in \mathbb{R}^{(\omega-1) \times \omega}$ is defined as the first-differencing matrix: 
	
	$$
	\boldsymbol{D}_{1} = 
	\begin{pmatrix}
		-1 		&  1 		& 0 		& 0 		& \dots 	& 0 \\
		0 		& -1 		& 1 		& 0 		& \dots 	& 0 \\
		\vdots & \ddots 	& \ddots	& \ddots	& \ddots	& \vdots \\
		0 		& \dots 	& 0		 	& -1 		& 1 		& 0 \\
		0 		& \dots 	& \dots 	& 0 		& -1 		& 1 \\
	\end{pmatrix}.
	$$
	
	Let $\boldsymbol{M} \in \mathbb{R}^{\omega \times L}$ be a matrix with $\omega$ age-specific log mortality rates (here single year ages $0$ to $99$) and $L$ different mortality schedules (e.g.~different countries and years). Then, the vector $\boldsymbol{c}_{1} \in \mathbb{R}^{(\omega-1)}$ contains the mean first differences of $\boldsymbol{M}$, i.e.~$\boldsymbol{c}_{1} = \frac{1}{L} \cdot \boldsymbol{D}_{1} \boldsymbol{M} \mathbf{1}_{L}$. The matrix $\hat{\boldsymbol{V}}_{1} \in \mathbb{R}^{(\omega-1) \times (\omega-1)}$ is the empirical covariance matrix constructed from the residuals calculated on $\boldsymbol{M}$. In other words, $\hat{\boldsymbol{V}}_{1}$ is the covariance matrix of $ \left( \boldsymbol{D}_{1} \boldsymbol{M} - \boldsymbol{c}_{1} \mathbf{1}_{\left(\omega - 1\right)}^{T} \right)^{T}$. 
	\textcite{Schmertmann2021DR} uses data from the Human Mortality Database \parencite{HMD20220711} to construct the residuals, i.e.~the columns of $\boldsymbol{M}$ contain all (complete) mortality schedules in the HMD.
	
	For the D-2 estimator, the matrix $\boldsymbol{D}_{2} \in \mathbb{R}^{(\omega-2) \times \omega}$ is the second differencing matrix:
	
	$$
	\boldsymbol{D}_{2} =
	\begin{pmatrix}
		1 		& -2 		& 1 		& 0 		& \dots 	& \dots 	& 0 \\
		0 		& 1 		& -2 		& 1 		& 0 		& \dots 	& 0 \\
		\vdots 	& \ddots 	& \ddots    & \ddots    & \ddots    & \ddots    & 0 \\
		0		& \dots     & 0			& 1			& -2		& 1			& 0 \\
		0 		& \dots 	& \dots		& 0			& 1 		& -2 		& 1 \\					
	\end{pmatrix}.
	$$
	
	Similar to the D-1 estimator but based on second instead of first differences, the vector $\boldsymbol{c}_{2} \in \mathbb{R}^{(\omega-2)}$ contains the mean second differences of a large database of age-specific log mortality rates $\boldsymbol{M}$. The covariance matrix $\hat{\boldsymbol{V}}_{2}$ is calculated analogously to the D-1 estimator but using $\boldsymbol{D}_{2}, \boldsymbol{c}_{2},$ and $\boldsymbol{1}_{\left( \omega - 2 \right)}$ instead.
	
	The D-LC estimator is inspired by the Lee-Carter model \parencite{LeeCarter1992JotASA}. Let $\boldsymbol{a} \in \mathbb{R}^{\omega}$ be the baseline mortality containing the rowmeans of $\boldsymbol{M}$. Further, let $\boldsymbol{X} \in \mathbb{R}^{\omega \times L}$ be the matrix of centered log mortality rates of $\boldsymbol{M}$. The age-specific deviations $\boldsymbol{b} \in \mathbb{R}^{\omega}$ are obtained via a singular-value decomposition on $\boldsymbol{X}$. Key is to re-arrange the Lee-Carter model $\log \left( \boldsymbol{m} \right) = \boldsymbol{a} + k \cdot \boldsymbol{b} $ to $ \log \left( \boldsymbol{m} \right) - \boldsymbol{a} = k \cdot \boldsymbol{b} $, where $k$ is a scalar. Then, the residuals can be calculated as \parencite{Schmertmann2021DR}
	$$
	\varepsilon = 
	\left( \boldsymbol{I}_{\omega} - \boldsymbol{b} \left( \boldsymbol{b}^{T}\boldsymbol{b} \right)^{-1} \boldsymbol{b}^{T} \right) \left( \log \boldsymbol{m} - \boldsymbol{a} \right) = \boldsymbol{D}_{3} \left( \log \boldsymbol{m} - \boldsymbol{a} \right),
	$$
	where $\boldsymbol{I}_{\omega}$ denotes the $\omega$-dimensional identity matrix. It follows that $\boldsymbol{c}_{3} = \boldsymbol{D}_{3} \boldsymbol{a} \in \mathbb{R}^{\omega}.$ The empirical covariance matrix $\hat{\boldsymbol{V}}_{3}$ is calculated analogously to that of the D-1 and D-2 estimators but using $\boldsymbol{D}_{3}, \boldsymbol{c}_{3},$ and $\boldsymbol{1}_{ \omega }$.
	
	For D-splines the same assumptions as in TOPALS hold regarding age-grouped input, i.e.~$D_g \sim \text{Poisson} \left( m_g N_g \right)$ and age-grouped mortality rates are the average of the single year mortality rates in the corresponding ages $x$ \parencite{Schmertmann2021DsplinesWithAgeGroupData}. \textcite[Appendix A-5]{Schmertmann2021DR} outlines how to obtain an approximate covariance matrix for the estimated log mortality rates which can be used to calculate pointwise confidence intervals.

	\section*{\normalfont \textit{SVD-model}}
	
	In the Bayesian hierarchical model based on principal components, the log mortality rates in age group $g = 1, \dots G$, area $a = 1, \dots, A$ and at time
	$t = 1, \dots, T$, $\log m_{g,a,t}$ are modeled as a linear combination of three principal components, $V_{p}, p = 1, \dots, 3, $ and a random error term $u_{g,a,t}$ that accounts for overdispersion \parencite{AlexanderEtAl2017D}: 
	$$
	\log m_{g,a,t} = \beta_{1,a,t} \cdot V_{1,g} + \beta_{2,a,t} \cdot V_{2,g} + \beta_{3,a,t} \cdot V_{3,g} + u_{g,a,t},
	$$
	where $\beta_{p,a,t}$ are the regression coefficients for the $p$-th principal component and $V_{p,g}$ is the age-specific value of $p$-th principal component.
	
	Let $\boldsymbol{X} \in \mathbb{R}^{F \times G}$ be a matrix that contains age-specific log mortality rates ($G$ columns) over time and space ($F$ rows). The $G$-dimensional vectors $\boldsymbol{V}_{p}$ are constructed by decomposing $\boldsymbol{X}$ by means of a singular-value decomposition, $\boldsymbol{X} = \boldsymbol{U}\boldsymbol{D}\boldsymbol{V}^{T}$. The columns of matrix $\boldsymbol{V}$ are the right-singular vectors.
	
	The model includes components to smooth over space and time. Geographical pooling is achieved by assuming that the regression coefficients $\beta_{p,a,t}$ are distributed around a common mean
	$$
	\beta_{p,a,t} \sim \mathcal{N} \left( \mu_{\beta_{p,t}}, \sigma_{\beta_{p,t}} \right),
	$$
	where $\mathcal{N} \left( \mu, \sigma \right)$ denotes the normal distribution with mean $\mu$ and standard deviation $\sigma$.
	
	Smoothing over time is achieved by assuming that the common mean $\mu_{\beta_{p,t}}$ is centered around the second differences 
	$$
	\mu_{\beta_{p,t}} \sim \mathcal{N} \left( 2 \cdot \mu_{\beta_{p,t-1}}
	- \mu_{\beta_{p,t-2}}, \sigma_{\mu_{\beta_{p,t}}} \right). 
	$$
	
	The model by \textcite{AlexanderEtAl2017D} estimates age-grouped mortality schedules for different subpopulations separately. \textcite{DharamshiEtAl2021PAA} develop this model further to simultaneously estimate mortality schedules for different subpopulations (e.g.~males and females, groups by socioeconomic status).

	\section*{\normalfont \textit{Comparative assessment of methods}}
	
	Table \ref{tab: overview_methods} shows an overview of TOPALS regression models, D-splines, and SVD-models with regard to their underlying statistical philosophy (frequentist/Bayesian), data requirements as well as the main generated output. In the following section, we discuss these characteristics and some of their implications. 
	
	The input data requirements for each method can be summarized as age-specific death counts and exposures (either age-grouped or single years) as well as some sort of incorporated demographic knowledge about human mortality schedules. The methods generally differ in whether mortality schedules for one or multiple areas/subpopulations\footnote{Methods which estimate mortality schedules separately may be used on areas or subpopulations or both.} are estimated at once. Regarding the number of mortality schedules that can be estimated at once, the SVD-model is designed to estimate mortality rates for multiple areas over time, relating them to each other by smoothing over space and time. The Bayesian TOPALS regression model by \textcite{RauSchmertmann2020DAO} estimates schedules for males and females simultaneously and the SVD-model for multiple subgroups correlates mortality schedules of different subpopulations with each other \parencite{DharamshiEtAl2021PAA}. Besides input data in terms of death counts and exposures, demographic knowledge is incorporated into each of the methods. However, the methods differ in the amount of demographic knowledge needed. As such, TOPALS regression models need only one mortality schedule to which the input data are related. In contrast, D-splines and SVD-models use multiple schedules.
	
	So far, we have discussed how the demographic knowledge is incorporated. However, we have not yet commented on what demographic knowledge is incorporated. Generally, for the estimation of mortality schedules at the subnational level with TOPALS regression models, the national mortality schedule of the corresponding country seems to be the first reasonable choice \parencites{GonzagaSchmertmann2016RBdEdP, SchmertmannGonzaga2018D, RauSchmertmann2020DAO}. For D-splines, \textcite{Schmertmann2021DR} constructs the residual vectors from all complete mortality schedules since $1970$ which are available in the HMD (in $10$-year intervals, excluding the country from which data are simulated). For SVD-models, the natural choice is the corresponding time series, e.g.~a national time series in the case of subnational regions as in \textcite{AlexanderEtAl2017D}. As such, \textcite{AlexanderEtAl2017D} use french national time series when estimating mortality schedules for french départements or the higher hierarchy such as the mortality schedules for U.S.~states when the level of interest are counties. 
	
	As for the output, all methods return either single year or age-grouped mortality rates. Comparing the input to the output data for each method, it is apparent that both TOPALS and D-splines always return single year mortality rates, even with age-grouped input. In contrast, the SVD-models return mortality rates on the same level of aggregation as the input data. Note that age-specific mortality rates are just one type of output generated. Besides the estimated schedules, other output might include measures of model fit (e.g.~AIC, BIC) or algorithmic fit diagnostics (e.g.~$\hat{R}$ in Bayesian models) or uncertainty intervals\footnote{We use the term \textit{uncertainty intervals} to include both frequentist \textit{confidence intervals} as well as Bayesian \textit{credible intervals}.}. Lastly, the SVD-models which include a time component allow for naturally studying changes in mortality over time and differences between areas by making use of estimated variance parameters as outlined in \textcite{AlexanderEtAl2017D}.  
	
	Finally, one concern with death counts is the possibility of overdispersion, i.e.~that the variance is larger than the expected value. TOPALS regression models may account for overdispersion as described by \textcite{SchmertmannGonzaga2018D}. While the model in the main publication does not account for overdispersion, the authors tested the model using a negative binomial instead of a Poisson distribution \parencite{SchmertmannGonzaga2018D}. D-splines do not account for overdispersion and this topic is not touched upon in the original publication \parencite{Schmertmann2021DR}. Finally, the SVD-model accounts for overdispersion by adding a random error term \parencites{AlexanderEtAl2017D}.
	
	\begin{table}[h]
		\small
		\renewcommand*{\arraystretch}{1.25} 
		\centering
		\caption[Overview of methods]{Overview of methods with regard to data requirements and output.}
		\label{tab: overview_methods}
		\begin{tabular}{lccccc}
			\hline
			\textbf{Method} 		& \textbf{Type} 	  &  	 \multicolumn{3}{c}{\textbf{Data Requirements}} 								& \textbf{Output}	\\
			&					  &	Input	 				 		 & Fitting             	& Standard 			        & 	\\
			\hline
			TOPALS 					& frequentist         & $D_x, N_x$ 			 		 	 & one area/subpopulation,			& one schedule $m_x$  		& $m_x$  \\
			\tiny \textcite{GonzagaSchmertmann2016RBdEdP} &     &						 & one point in time	& 							&  \\
			TOPALS 				   	& Bayesian            & $D_x, N_x$ 			 		 	 & one area/subpopulation,			& one schedule $m_x$				& $m_x$ \\
			\tiny \textcite{SchmertmannGonzaga2018D} &     	  &									 & one point in time						& 							& \\
			TOPALS			 		& Bayesian 			  & $D_g, N_g$ 			 		 	 & multiple areas, 		& one schedule $m_x$ 		& $m_x$ \\
			\tiny \textcite{RauSchmertmann2020DAO} &     		  &									 & one point in time,	& 							& \\		
			& & & two sexes & &  \\
			TOPALS 					& frequentist         & $D_g, N_g$ 			 		 	 & one area/subpopulation,			& one schedule $m_x$  		& $m_x$ \\
			\tiny \textcite{Schmertmann2022TopalsFittingWithGroupedData} &  	 &  				 & one point in time	& 							& \\
			D-splines			 	& frequentist 		  & $D_x, N_x$ 			 		 	 & one area/subpopulation,			& mult.~schedules $m_x$  	& $m_x$ \\
			\tiny \textcite{Schmertmann2021DR} &     			  &									 & one point in time	& 							& \\	
			D-splines			 	& frequentist 		  & $D_g, N_g$ 			 		 	 & one area/subpopulation,			& mult.~schedules $m_x$  	& $m_x$ \\
			\tiny \textcite{Schmertmann2021DsplinesWithAgeGroupData} &   &						 & one point in time	& 							& \\		
			SVD-model				& Bayesian 			  & $D_g, N_g$ 			 		 	 & multiple areas, 		& mult.~schedules $m_g$ 	& $m_g$ \\
			\tiny \textcite{AlexanderEtAl2017D} &     		  &									 & time series			& 							& \\	
			SVD-model				& Bayesian 			  & $D_g, N_g$ 			 		 	 & multiple areas, 		& mult.~schedules $m_g$ 	& $m_g$ \\
			\tiny \textcite{DharamshiEtAl2021PAA} &     		  &									 & time series,			& 							& \\
			&     		  &									 & multiple subgroups	& 							& \\		
			\hline	
			\multicolumn{6}{l}{
				\begin{tabularx}{0.95\textwidth}{X}
					\textit{Note:} $D$ and $N$ are death counts and exposures, $m$ is the mortality rate. Index $x$ is for single year ages while $g$ stands for age-grouped data. For example, $D_x$ are death counts for single year ages and $D_g$ age-grouped death counts (usually 5-year intervals).	
				\end{tabularx}
			} 
		\end{tabular}
	\end{table}

	\section{Simulation Study}
	\label{sec: simulation study ADEMP}
	
	We now describe the set-up of the simulation study following the ADEMP structure proposed by \textcite{MorrisEtAl2019SiM}. ADEMP is a framework for writing protocols for simulation studies and stands for (i) Aims, (ii) Data-generating mechanisms, (iii) Estimands (target of analysis), (iv) Methods, and (v) Performance measures \parencite{MorrisEtAl2019SiM}. Table \ref{tab: overview_protocol_sim_stud} shows the outline of our simulation study with respect to the ADEMP structure. This section and especially Table \ref{tab: overview_protocol_sim_stud} follow the protocol by \textcite{KiprutoSauerbrei2022PO}. We also add a section on the software used.
	
	\begin{table}
		\renewcommand*{\arraystretch}{1.25} 
		\small
		\centering
		\caption[Overview of simulation study]{Overview of simulation study following the ADEMP structure.}
		\label{tab: overview_protocol_sim_stud}
		\begin{tabularx}{\textwidth}{lX}
			\hline
			Aims 	 												& - To evaluate the ability to estimate age-specific log mortality rates. \\
			& - To evaluate the ability to estimate life expectancy. \\
			& - To study the sensitivity to (a) exposure size and (b) the incorporated demographic knowledge. \\
			\hline
			Data Generating 										& \textbf{General:} \\
			Mechanism													&  One dataset consists of age-specific death counts, exposures (and thus mortality rates)
			for {$20$ regions/subpopulations} and one sex (here: males) over $11$ years ($1995 - 2005$), i.e. $20 \times 11 = 220$ mortality schedules. \\
			& \textbf{Fixed parameters:} \\
			& - $r$: reference population (Germany, HMD code: DEUTNP) \\
			& - $i = 1, \dots, 20$: $20$ regions  \\
			& - $t = 1995, 1996, \dots, 2004, 2005$: $11$ years \\
			& - $x = 0, 1, \dots, \left( \omega - 1 \right)$: single year ages (No open age interval from Step 3 onwards, $\left( \omega - 1 \right) = 99$). \\
			& - $a_i \sim \mathcal{U} \left( -0.75, 0.75 \right) $ and $b_i \sim \mathcal{U} \left( 0.7, 1.3 \right)$ (as in \textcite{AlexanderEtAl2017D}). $a_i$ and $b_i$ are fixed over time. \\
			& \textbf{Procedure:} \\
			& 1. $D_x^{tr}$ and $N_x^{tr}$ are yearly age-specific death counts and exposures from a reference population $x = 0, 1, \dots, 100,$ where the last age group is open. Smooth for each year separately (\texttt{R}-package \texttt{MortalitySmooth 2.3.4} \parencite{Camarda2012JoSS}). \\
			& 2. Generate $20$ regions/subpopulations using the Brass model (as in \textcite{AlexanderEtAl2017D}): \\
			& (i) $Y_x^{it} = a_{i} + b_{i} Y_x^{tr}$, where $Y_x = - \frac{1}{2} \log \left( \frac{l_x}{1 - l_x} \right)$. \\
			& (ii) Convert $Y_x^{it}$ to age-specific mortality rates $m_x^{it}$. Keep all ages $x \leq 99.$ \\
			& 3. Simulate death counts for each $i$ and $t$. The \textit{true} mortality rate for TOPALS/D-splines is for single year ages, $m_x^{it}$, while that of the SVD-model is for age-grouped data, $m_g^{it}$. For that reason, we use different procedures for simulating the data. \\
			& \begin{tabular}[t]{l|l}
				\underline{TOPALS/D-splines:} & \underline{SVD-model:} \\
				$D_x^{\text{sim},it} \sim \text{Poisson} \left( m_x^{it} N_x^{t} \right),$ & (i) Calculate age specific death counts: $D_x^{it} = m_{x}^{it}N_x^{t}$ \\
				where $N_{x}^{t}$ is age-specific & (ii) Aggregate: $D_g^{it} = \sum_{x \in X_g} D_x^{it} $, $N_g^{t} = \sum_{x \in X_g} N_x^{t} $. \\
				exposure in year $t$ & (iii) Calculate the \enquote{true} mortality rates: $m_g^{it} = \frac{D_g^{it}}{N_g^{t}}$ \\
				& (iv) Simulate death counts: $D_g^{\text{sim},it} \sim \text{Poisson} \left( m_g^{it}N_g^{t} \right)$ \\
			\end{tabular} \\ 
			& \textbf{Exposure sizes:} \\
			& $N^{t} \in \left\{ 1\,000, 5\,000, 10\,000, 25\,000, 50\,000, 75\,000, 100\,000, 1\,000\,000 \right\}$ \\
			& \textbf{Simulation runs:} \\
			& We simulate $1\,000$ datasets for each of the $ 8 $ exposure sizes ($n_{\text{sim}} = 1\,000 $). \\
			\hline
			Target of Analysis										& Age-specific log mortality rates ($\log \left( m_x \right)$, $\log \left( m_g \right)$) and life expectancy ($e_0$) in the year $2\,000$. \\
			\hline
			Methods												   	& \begin{tabular}[t]{ll}
				\hline
				\textbf{Method} & \textbf{Parameters} \\
				\hline
				TOPALS & demographic knowledge, input data (single year, age-grouped) \\
				D-splines & demographic knowledge, input data (single year, age-grouped), estimator \\
				SVD-model & demographic knowledge \\
				$e_0$ from raw data & - \\
				\hline
			\end{tabular} \vspace*{0.3cm} \\
			\hline
			Performance & bias, root mean squared error, empirical standard error, coverage of uncertainty \\
			Measures		& intervals, width of uncertainty intervals (as outlined in \textcite{MorrisEtAl2019SiM})  \\		
			\hline
			\multicolumn{2}{l}{
				\begin{tabularx}{0.8\textwidth}{X}
					\textit{Note:} The idea and structure of the table is as in \textcite[Table 1]{KiprutoSauerbrei2022PO}.
				\end{tabularx}
			} 
		\end{tabularx}
	\end{table}
	
	\section*{\normalfont \textit{Aims}} \label{subsubsec: aims}
	
	The aims of our simulation study are (i) to evaluate the ability of each method to estimate age-specific mortality rates and (ii) life expectancy, and (iii) to study the sensitivity of each method to (a) exposure size and (b) the incorporated demographic knowledge.
	
	\section*{\normalfont \textit{Data-generating process}} \label{subsubsec: DGP}
	
	We assume that age-specific male death counts in region $i$ and year $t$, $D_x^{it}$, are Poisson distributed with age-year-region-specific mortality rate $m_x^{it}$ and age-year-specific exposure size $N_x^{t}$, i.e.~$D_{x}^{it} \sim \text{Poisson} \left( m_x^{it} N_x^{t} \right), i = 1, \dots, 20, t = 1995, \dots, 2005$. We define one dataset as the age-specific death counts and exposures (and thus age-specific mortality rates) for all $20$ subpopulations/regions over $11$ years. In other words, we can picture one dataset as the age-specific deaths and their corresponding exposures for males in one country with $20$ regions/subpopulations over $11$ years. 
	
	We obtain the age-region-year specific mortality rates, $m_{x}^{it}$ by means of a Brass relation model \parencite{Brass1971BAD} as in \textcite{AlexanderEtAl2017D}\footnote{The procedure in \textcite{AlexanderEtAl2017D} is not described in sufficient detail in order to reconstruct the exact simulation design.}. In the Brass relational model, the age-specific logits of the survivorship, $Y_x^{it}$ are linearly related to the logits of the survivorship in a reference population $r$, $Y_x^{tr}$:
	$$
	Y_x^{it} = a_i + b_i Y_x^{tr},
	$$
	where $ Y_x^{it} = \text{logit}  \left(1 - l_x^{it} \right) $ and $Y_x^{tr} = \text{logit}  \left(1 - l_x^{tr} \right)$ with $l_x^{it}, l_x^{tr}$ as the proportions of survivors until age $x$ (radix $l_0 = 1$) \parencite{Brass1971BAD}. The logit is defined as $\text{logit} \left( p \right) = \frac{1}{2} \log \left( \frac{p}{1-p} \right)$ \parencite{Brass1971BAD}.\footnote{The \enquote{standard} logit known in statistics is usually defined without the scalar $ \frac{1}{2}$. \textcite[74]{Brass1971BAD}, however, explicitly refers to the logit with scalar and its use in bio-assays.}
	
	In order to simulate age-specific mortality rates for $20$ regions over $11$ years, we need to select a reference population and Brass parameter vectors $\mathbf{a} \in \mathbb{R}^{20}$ and $\mathbf{b} \in \mathbb{R}^{20}$. Following \textcite{AlexanderEtAl2017D} we randomly draw the Brass parameters from uniform distributions, $a_i \sim \mathcal{U} \left( -0.75, 0.75 \right) $ and $b_i \sim \mathcal{U} \left( 0.7, 1.3 \right), i = 1, \dots, 20.$ For the entire simulation, we repeat this procedure once. Thus, the $a_i$ and $b_i$ are fixed over time. This ensures that the relationship between the reference population and the corresponding regions is stable over time. As reference population we choose German data. Before applying the Brass model we smooth the mortality rates of the German reference population using the \texttt{R}-package \texttt{MortalitySmooth} \parencite{Camarda2012JoSS}. Death counts and exposures are obtained from the Human Mortality Database \parencite{HMD20220711}\footnote{Death counts and exposures downloaded on 2022-07-11.}. After generating $Y_x^{it}$, we convert these values into age-specific mortality rates. Section \ref{sec: sm_artifical_regions} in the supplementary material displays the Brass parameters (Table~\ref{tab: sm_brass_parameters}) and plots of life expectancies as well as the human mortality schedules for the $20$ regions over time (Figures~\ref{fig: sm_e0_cutoff_true_subpopulations}--\ref{fig: sm_logmx_over_time_by_year_subpopulations}).
	
	We simulate death counts for each $i$ and $t$ assuming that the number of deaths is Poisson distributed. While the procedure to simulate death counts is the same for TOPALS and D-splines, it is different for the SVD-model. The reason for using different procedures is that D-splines and TOPALS estimate single year mortality rates while the SVD-model estimates age-grouped mortality rates (see Table \ref{tab: overview_methods}). Thus, for D-splines and TOPALS the \textit{true} underlying mortality rates need to be for single year ages and those for the SVD-model need to be age-grouped.
	
	For D-splines and TOPALS, we simulate death counts for each $i$ and $t$ according to $D_x^{\text{sim},it} \sim \text{Poisson} \left( m_x^{it} N_x^{t} \right)$, where $N_x^{t}$ are the age-specific exposures in year $t$. Similar to \textcite{Schmertmann2021DR}, the age-specific exposures in year $t$, $N_x^{t}$, are re-weighted according to the age structure in the German data for year $t$. In other words, if the desired exposure size in year $t$ is  $ N^{t} = \sum_{x} N_x^{t} = 10\,000$, the age-specific value is obtained by multiplying the proportion of person-years lived in each age group in the reference population by the desired number of person-years, i.e. $ N_x^{t} = \frac{N_x^{tr}}{N^{tr}} \cdot N^{t} $. Thus, each region/subpopulation has the same age structure in a specific year as all data are simulated from the German reference population. Following \textcite{Schmertmann2021DR}, we ensure that there are at least two deaths in each subpopulation. 
	
	For the SVD-model we simulate age-grouped death counts according to $D_g^{\text{sim},it} \sim \text{Poisson} \left( m_g^{it}N_g^{t} \right)$. We obtain the \textit{true} age-grouped mortality rates $m_g^{it}$ and exposures $N_g^{t}$ as follows. First, we calculate single year death counts, $D_x^{it} = m_{x}^{it}N_x^{t}$. Second, we aggregate single year death counts and exposures into age groups, $D_g^{it} = \sum_{x \in X_g} D_x^{it} $ and $N_g^{t} = \sum_{x \in X_g} N_x^{t} $. Third, we calculate the \enquote{true} age-grouped mortality rates: $m_g^{it} = \frac{D_g^{it}}{N_g^{t}}$. Fourth, we simulate age-grouped death counts: $D_g^{\text{sim},it} \sim \text{Poisson} \left( m_g^{it}N_g^{t} \right)$. As for TOPALS and D-splines the exposures $N_g^{t}$ are re-weighted according to the age-structure in the reference population.
	
	We repeat the aforementioned procedure for eight different exposure sizes, 
	$$ N^{t} \in \left\{ 1\,000, 5\,000, 10\,000, 25\,000, 50\,000, 75\,000, 100\,000, 1\,000\,000 \right\}.$$ 
	For each of these exposure sizes, we simulate $1\,000$ datasets.
	
	\section*{\normalfont \textit{Target of analysis}} \label{subsubsec: Target of Analysis}
	
	We are interested in age-specific log mortality rates ($\log \left( m_x \right) $ or $\log \left( m_g \right)$) and life expectancy ($e_0$). On a technical note, we use the term life expectancy to refer to partial life expectancy from the ages $0-99$ (there is no open age interval). This approach has previously been used in the literature (e.g.~\textcite{Schmertmann2021DR}). Informal experiments showed that partial life expectancy (ages $0-99$) is mostly very similar to life expectancy at birth. However, differences can occur when many people are alive at the oldest ages, i.e.~when the (estimated) schedules are implausible given mortality rates in contemporary human populations.
	
	\section*{\normalfont \textit{Methods}} \label{subsubsec: methods}
	
	We evaluate frequentist TOPALS regression, D-splines, and an SVD-model. These methods produce estimates of age-specific log mortality rates from which we calculate life expectancy. As outlined in Table \ref{tab: overview_protocol_sim_stud}, each of the three methods comes with a set of parameters that needs to be specified by the user. First, all three methods incorporate demographic knowledge. Second, TOPALS and D-splines can be used with single year or age-grouped input data. Third, three different estimators, namely D1, D2, and DLC, exist for D-splines. It is an important part of the estimation process to specify these parameters.
	
	One of the main features of a \textit{good} method is its ability to produce accurate estimates under a range of different settings (robustness of a method). This robustness can refer to a method's parameters (e.g.~type of demographic knowledge) or dataset characteristics (e.g.~exposure size). In addition to testing the methods with different exposure sizes (see earlier Section on Data-generating process), we vary the aforementioned parameters of each method (incorporated demographic knowledge, single-year and age grouped input for TOPALS and D-splines, all three D-spline estimators). For each method, we test three different types of demographic knowledge.
	
	It would be desirable to test the methods with respect to the incorporated demographic knowledge in a way that is comparable across methods. However, as outlined in Section \ref{sec: methods} the methods generally differ in the amount of data needed as well as the way the demographic knowledge is constructed and incorporated, making \textit{fair} evaluations hardly possible. Therefore, we construct demographic knowledge (dk) following some general principles:
	
	\begin{enumerate}
		\item Canonical choice (dk1). 
		\begin{itemize}
			\item TOPALS: The national schedule. Smoothed male national schedule of Germany for the year $2000$ (HMD code: \enquote{DEUTNP}).
			\item D-splines: All complete HMD mortality schedules from $1970-2019$ excluding Germany (HMD codes: \enquote{DEUTNP}, \enquote{DEUTW}, \enquote{DEUTE}), largely as in \textcite{Schmertmann2021DR}.
			\item SVD-model: Smoothed national time series ($1995 - 2005$) of the reference population (Germany, DEUTNP).
		\end{itemize}	
		\item Wrong/unintuitive (dk2).
		\begin{itemize}
			\item TOPALS: Smoothed mortality schedule from the same sex (male) but a different country with similar life expectancy: Smoothed male national schedule of France for the year $2000$ (HMD code: \enquote{FRATNP}).
			\item D-splines: Historic HMD data only, i.e.~all complete schedules before $1970$ excluding Germany (HMD codes: \enquote{DEUTNP}, \enquote{DEUTW}, \enquote{DEUTE}).
			\item SVD-model: All complete mortality schedules from $1970-2018$ excluding Germany (HMD codes: \enquote{DEUTNP}, \enquote{DEUTW}, \enquote{DEUTE}). 
		\end{itemize}
		\item Wrong sex (dk3). 
		\begin{itemize}
			\item TOPALS: Smoothed mortality schedule from the other sex (female) and a different country but with similar life expectancy: Smoothed female national schedule of France for the year $1965$ (HMD code: \enquote{FRATNP}).
			\item D-splines \& SVD: Construct as dk1 but from female, instead of male data.
		\end{itemize}
	\end{enumerate}
	
	We use death counts and exposures for single year ages from the Human Mortality Database \parencite{HMD20220711} to calculate mortality rates and construct the demographic knowledge. For TOPALS and the SVD-model we use yearly death counts and exposures which we aggregate into age groups for the SVD-model. For D-splines, as in \textcite{Schmertmann2021DR} we use single year aged data for 10-year intervals and smooth the precalibrated vector $c_i, i = 1, 2, 3$.
	
	While the three types of demographic knowledge follow the same principles, comparisons across methods are difficult and especially for dk2 unintuitive. As such, we expect that TOPALS performs similar for dk1 and dk2 as the levels of mortality are similar. For D-splines, however, we use outdated, historic data. Lastly, we exclude the reference population from the construction of demographic knowledge whenever possible to avoid overoptimism: We used the reference population to construct schedules for $20$ region/subpopulations from which we simulated data. Thus, we want to avoid feeding the same data twice (when simulating data and during the estimation process). For this reason, dk1 and dk2 are similar for TOPALS but dk2 does not use German data. For the same reason, we do not use the $20$ generated schedules from which we directly simulate data which---at first sight---seems like a good choice for the SVD-model. For TOPALS and the SVD model, the three types of demographic knowledge are shown in the supplementary material, Section \ref{sec: dem_know_topals_svd}, Figures \ref{fig: sm_dk1_TOPALS_males} -- \ref{fig: sm_plot_pcs_SVD_model}.
	
	\section*{\normalfont \textit{Performance measures}} \label{subsubsec: performance measures}
	
	We evaluate the methods using the following indicators: bias (bias), root mean squared error (RMSE), empirical standard error (empSE) as well as coverage and width of the uncertainty intervals (covCI and widCI) \parencite{MorrisEtAl2019SiM}. The formulas are shown in the supplementary material, Section \ref{sec: sm_performance_measures}. We calculate the performance measures from the (estimated) age-specific log mortality rates (as in e.g.~\textcite{Schmertmann2021DR}). We deliberately choose log mortality rates instead of mortality rates. Previous studies did not comment on this choice even though it influences the performance metrics (except for coverage) and the picture that is shaped by them. Taking the difference of the logarithm of two values eliminates the order of magnitude. In other words, more weight is put on differences in low mortality regimes, i.e.~at younger ages. We do not calculate the performance measures for life expectancy but instead visualize the results by means of boxplots.
	
	\section*{\normalfont \textit{Software}}
	\label{subsec: software}
	
	We conduct all simulations using \texttt{R} \parencite{R2022} (all simulations are run with version \texttt{4.2.2}, pre-processing and analysis with \texttt{4.2.1}). In addition, we use several add-on packages: \texttt{batchtools} \parencite{batchtools} to interact with a high-performance cluster, \texttt{checkmate} \parencite{checkmate} for checking code, \texttt{cowplot} \parencite{cowplot.1.1.1} to arrange several plots in one figure, \texttt{data.table} \parencite{data.table} for data wrangling, \texttt{ggplot2} \parencite{ggplot2} for plotting, \texttt{here} \parencite{here.1.0.1} to handle paths, \texttt{LifeTable} \parencite{LifeTable.2.0.2} for life table calculations, \texttt{MortalitySmooth} \parencite{Camarda2012JoSS} to smooth mortality schedules, \texttt{rstan} \parencite{rstan.2.21.8} for the SVD-model, and \texttt{xtable} \parencite{xtable} to create \LaTeX~tables.
	For frequentist TOPALS regression and D-splines we used the \texttt{R} code available in \textcite{Schmertmann2022TopalsFittingWithGroupedData} and \textcite{Schmertmann2021DsplinesWithAgeGroupData}. For the SVD-model Ameer Dharamshi shared \texttt{stan} code with us (A.~Dharamshi, personal communication, May 2022). The specification of the SVD-model is shown in the supplementary material, Section \ref{sec: sm_svd_model_formulation}. Full code to reproduce the analysis is available at \url{https://github.com/estherden/sae_evaluation}.

	\section{Results}
	\label{sec: results} 
	
	\subsection{TOPALS \& D-splines}
	\label{subsec: results topals dsplines}
	
	We present results for three artificial areas/subpopulations (high, medium, low mortality rates) and exposures of $1\,000, 5\,000, 50\,000, 100\,000$, and $1\,000\,000$.
	Results shown are based on $1\,000$ simulation repetitions. At an exposure size of $1\,000\,000$ a number computations failed for D-splines with age-grouped input and dk2 ($988$ for D1, $2\,478$ for D2, and $2\,753$ for DLC distributed across all $20$ different regions/subpopulations). This is without consequences for the results shown in the main body of the paper but should not remain unmentioned.

	\subsection*{\normalfont \textit{Sensitivity to exposure size}}
	
	Figure \ref{fig: Topals_Dsplines_over_exposures_idab9_male} shows the performance measures for the age-specific log-mortality rates for different exposure sizes but one fixed region and incorporated demographic knowledge (canonical choice, dk1). Some general patterns emerge. First, bias, empirical standard error, RMSE, and the width of the confidence intervals decrease with increasing exposure. Second, the coverage of the confidence intervals decreases with increasing exposure. Third, the results for age-grouped and single year aged input are similar for each method with an exception for the coverage of the confidence intervals. For the highest ages and with increasing exposures the coverage with regard to the input data (age-grouped vs.~single years) differs. In these cases, the coverage for age-grouped input is lower. Figure \ref{fig: sm_Topals_Dsplines_over_exposures_idab9_male} in Section \ref{sec: sm_additional results} in the supplementary material shows similar results for the remaining exposure sizes.
	
	Instead of studying general patterns, we can also look at the performance of specific methods and the interplay of the performance measures. TOPALS, for example, is relatively unbiased for all ages and over all exposures. The empirical standard errors are highest for younger ages and decrease with increasing exposure size. This combination of (relatively) small bias but uncertainty in the estimates is reflected in the RMSE which follows the same general pattern as the empirical standard error. The coverage of the confidence intervals is higher than $95\%$ for the smallest exposures and relatively close to the desired level for exposures up to $100\,000$ (with an exception for the highest ages). For an exposure of $1\,000\,000$, however, coverage deteriorates, especially for higher ages and age-grouped input. The good coverage comes with wide confidence intervals, especially for younger ages.
	
	Coverage for the D1 and DLC estimator of D-splines is well below the desired level of $95\%$ which is also reflected in tight confidence intervals. Bias and coverage are especially low for younger ages. While the performance measures based on estimates by the D1 and DLC estimators are largely similar, a different picture emerges for the D2 estimator. Coverage is relatively good but comes with very wide confidence intervals. In fact, for very small exposures some implausible schedules were produced (not shown).

	\subsection*{\normalfont \textit{Sensitivity to incorporated demographic knowledge}}
	
	In Figure \ref{fig: Topals_Dsplines_over_demKnow_idab9_male} we vary the incorporated demographic knowledge while keeping the exposures size fixed at $50\,000$ and one region (same region as in Figure \ref{fig: Topals_Dsplines_over_exposures_idab9_male}). It is evident that all methods are (at least somewhat) sensitive to the incorporated demographic knowledge. For TOPALS, the standard schedules dk1 (German males in the year $2000$) and dk2 (French males in the year $2000$) are very similar. In this case, the trajectories of the performance measures are largely the same. For dk3 (french females in the year $1965$) larger biases can be observed at younger ages, especially before and after the accident hump.
	
	For D-splines dk1 and dk2 are constructed with different rationales: While dk1 follows \textcite{Schmertmann2021DR} using all complete male schedules in the HMD since $1970$, dk2 is based on historic data using all complete male schedules before $1970$. No large changes in bias can be observed for D1 and DLC. For D2 the bias fluctuates around zero for dk1 but increases sharply for dk2 and younger ages. Lastly, for dk3, all methods show larger bias around the age of $20$. This indicates an influence of the different shape of male and female schedules after the accident hump.
	
	For both TOPALS and D-splines the width of the confidence intervals is relatively stable over the three different types of incorporated demographic knowledge. The coverage, however, changes. This is especially prevalent for D-splines. For example, around the age of $20$ coverage is fairly high for dk1 and dk2 but drops considerably for dk3. Again, this is correlated with the presence of the accident hump in the underlying male data which is not so prevalent in the female data.

	\subsection*{\normalfont \textit{Performance for multiple regions}}
	
\begin{sidewaysfigure}
		\centering
		\includegraphics[width = \textwidth]{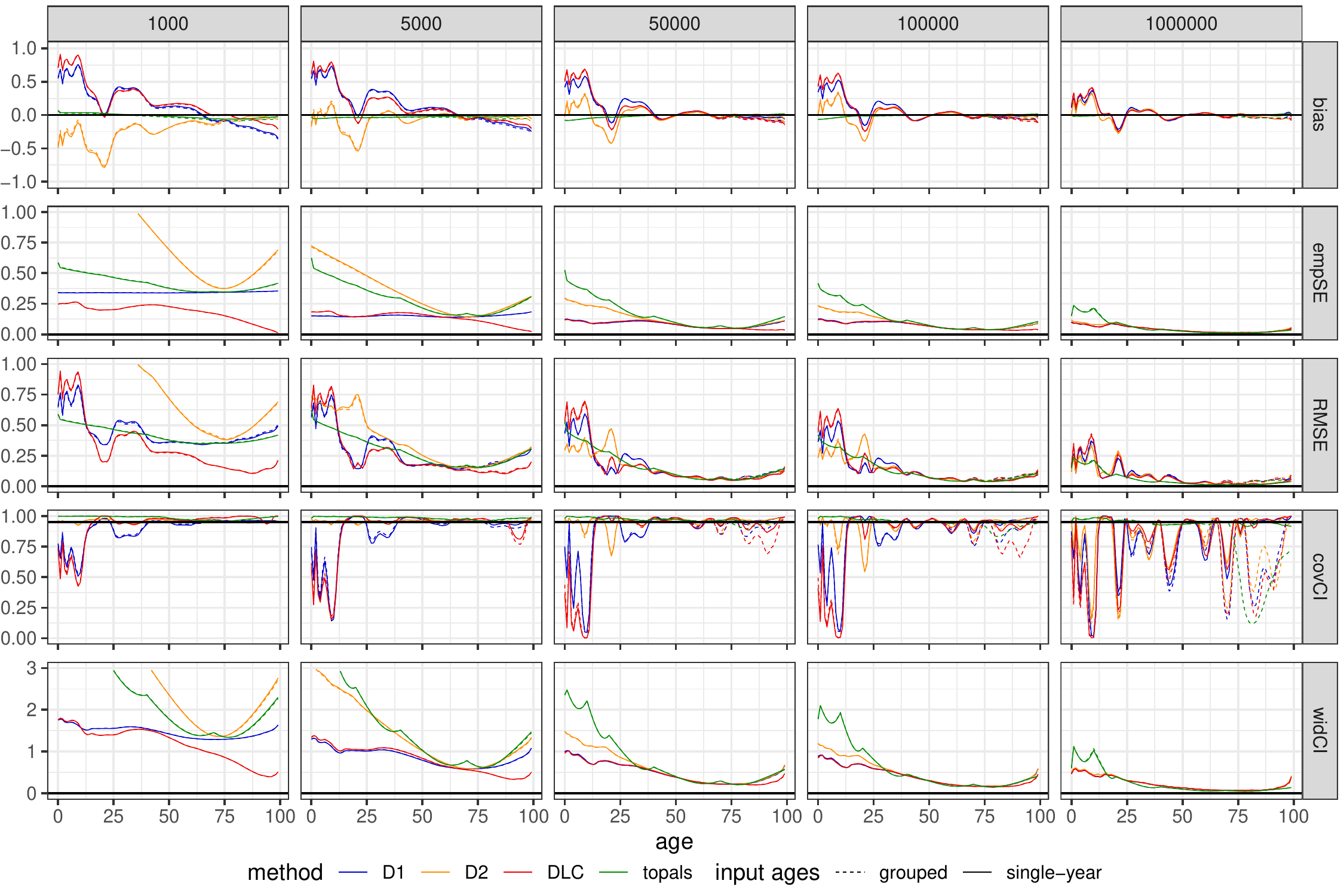}
		\caption{Performance measures for TOPALS and D-splines (D1, D2, DLC) over five different exposures sizes. Results shown for age-grouped (dashed line) and single year age (solid line) input data. Results shown are for the year $2000$ for one fixed region (similar to schedule of German males in the year $2\,000$ with Brass parameters $a = 0.018$ and $b = 1.035 $) and the canonical demographic knowledge (TOPALS: German males for the year $2000$. D-splines: All complete male mortality schedules in the HMD since $1970$.). The empirical standard error is not comparable across methods in the presence of bias \parencite{MorrisEtAl2019SiM}. Additionally, some lines are cut-off on the vertical axis due to very large values (e.g.~width of confidence intervals for D2 at small exposures). The idea of this figure is based on \textcite[Figure 1]{MorrisEtAl2019SiM}.}
		\label{fig: Topals_Dsplines_over_exposures_idab9_male}
	\end{sidewaysfigure}

\begin{sidewaysfigure}
		\centering
		\includegraphics[width = \textwidth]{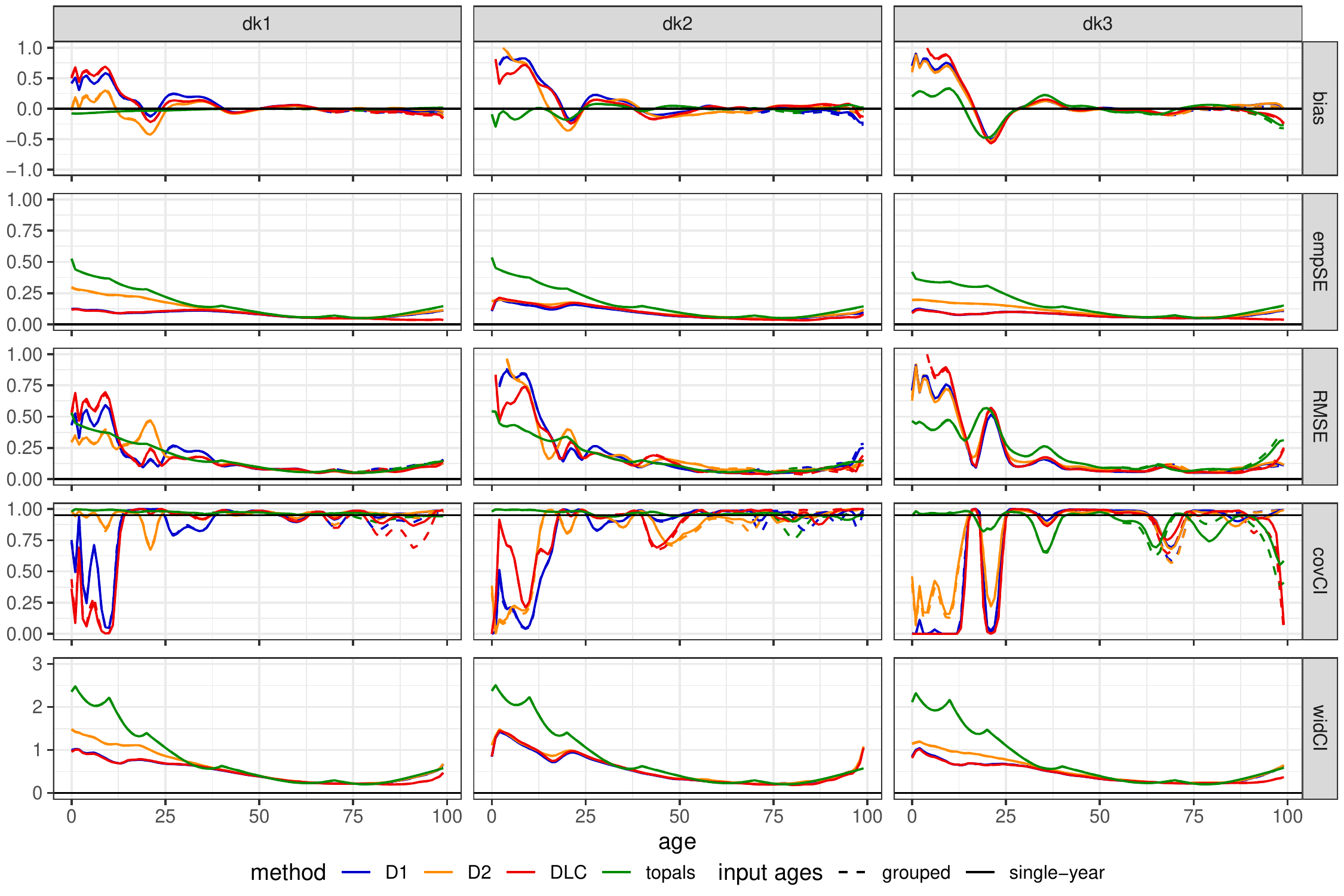}
		\caption{Performance measures for TOPALS and D-splines (D1, D2, DLC) with varying incorporated demographic knowledge but fixed exposure size ($50\,000$) and fixed region (same as in Figure \ref{fig: Topals_Dsplines_over_exposures_idab9_male} and similar to schedule of German males in the year $2\,000 $ with Brass parameters $a = 0.018$ and $b = 1.035 $). Figure based on results for the year $2000$. Comparisons across methods are not possible for dk2 as the rationale behind the standard differs (TOPALS: similar level of mortality as dk1. D-splines: historic data for dk2 but more recent for dk1). The empirical standard error is not comparable across methods in the presence of bias \parencite{MorrisEtAl2019SiM}. Some lines are cut-off on the vertical axis due to very large values (e.g.~bias for D2 at the youngest ages). The idea of this figure is based on \textcite[Figure 1]{MorrisEtAl2019SiM}.}
		\label{fig: Topals_Dsplines_over_demKnow_idab9_male}
	\end{sidewaysfigure}

\begin{sidewaysfigure}
		\centering
		\includegraphics[width = \textwidth]{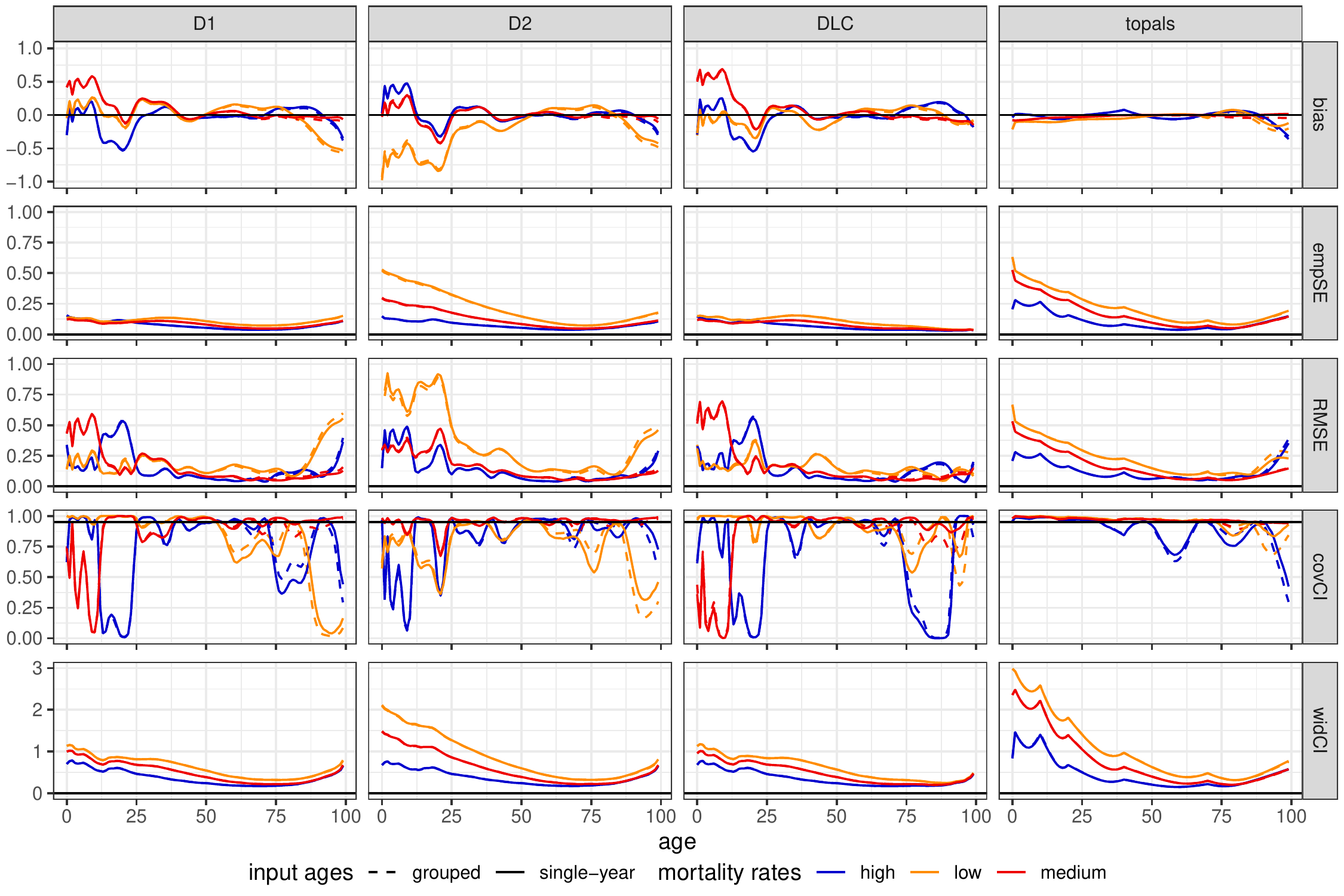}
		\caption{Performance measures for TOPALS and D-splines (D1, D2, DLC) for three different regions (low ($a = -0.659, b = 0.880$), medium ($a = 0.018$ and $b = 1.035 $), high ($a = 0.693, b = 0.850$) mortality setting) but fixed exposure size ($50\,000$) and incorporated demographic knowledge (dk1). Results shown for the year $2000$. The empirical standard error is not comparable across methods in the presence of bias \parencite{MorrisEtAl2019SiM}. The idea of this figure is based on \textcite[Figure 1]{MorrisEtAl2019SiM}.}
		\label{fig: Topals_Dsplines_over_methods_many_regions}
	\end{sidewaysfigure}

	Figures \ref{fig: Topals_Dsplines_over_exposures_idab9_male} and \ref{fig: Topals_Dsplines_over_demKnow_idab9_male} show results for one fixed region/subpopulation while varying exposure size and demographic knowledge. Lastly, we are also interested in understanding whether the observed patterns for each method are stable over different regions. Figure \ref{fig: Topals_Dsplines_over_methods_many_regions} indicates that this is not the case. While the general trajectories of the bias are similar over age there is considerable variability for D-splines, less so for TOPALS.
	The empirical standard error, however, varies more between regions for D2 and TOPALS than for D1 and DLC which is reflected in the RMSE. Coverage of the confidence intervals varies across regions for D-splines. For example, for D1 and DLC coverage is very low for young ages and medium mortality but generally higher for the high mortality region. For all methods the width of the confidence intervals is ordered with the low mortality setting having the widest and the high mortality setting having the tightest confidence intervals.
	
	\subsection*{\normalfont \textit{Ability to estimate life expectancy}}
	
\begin{sidewaysfigure}
		\centering
		\includegraphics[width = \textwidth]{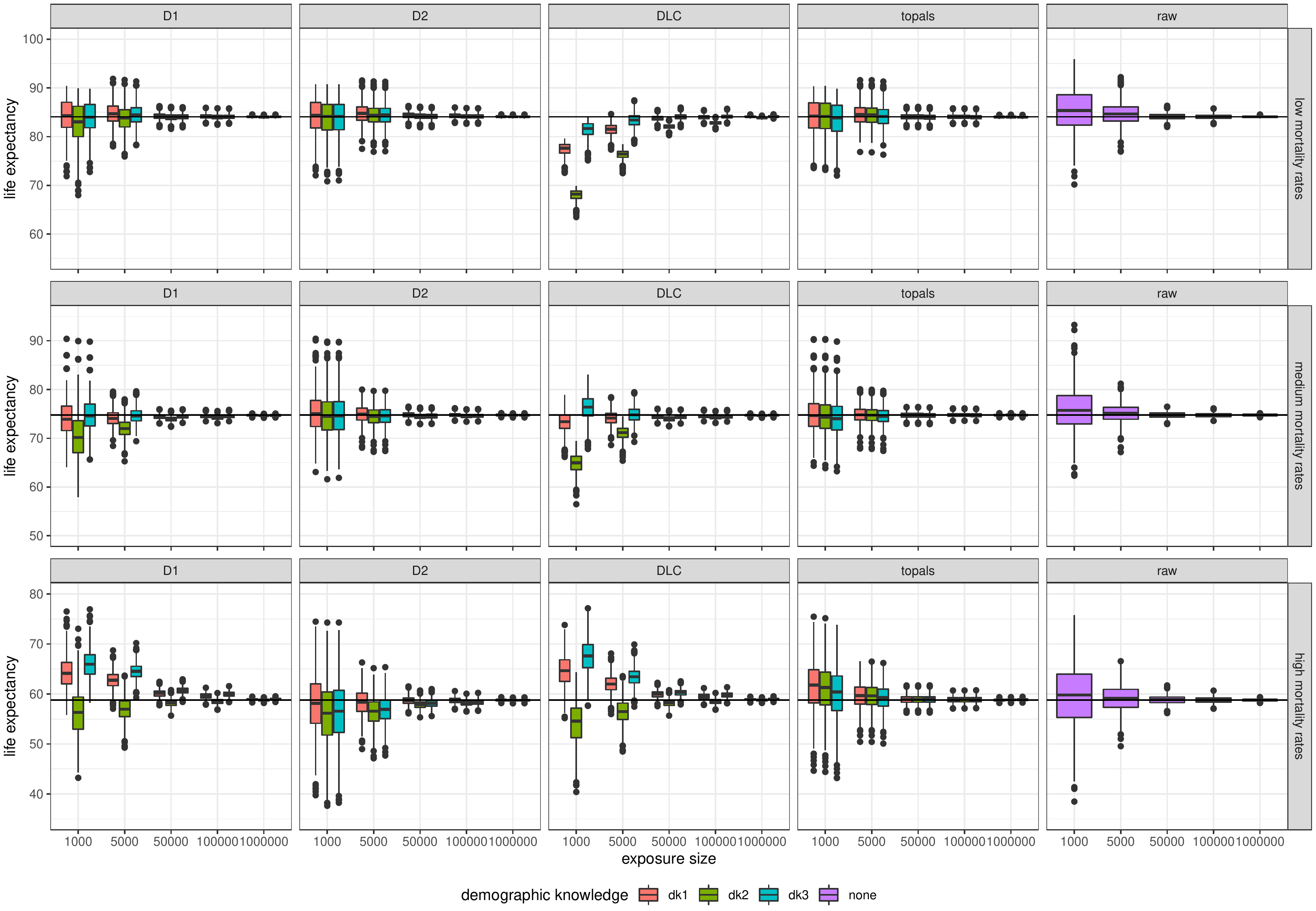}
		\caption{Boxplots of partial life expectancy (ages $0-99$) over all simulation runs for TOPALS, D-splines (D1, D2, DLC) for single year age input data for the year $2000$. Estimates from the raw data are shown as well. The horizontal line is the true life expectancy. Rows correspond to different regions/subpopulations with different levels of mortality (low ($a = -0.659, b = 0.880$), medium ($a = 0.018$ and $b = 1.035 $), high ($a = 0.693, b = 0.850$) mortality setting). Columns correspond to methods and colours to the incorporated demographic knowledge. Ranges of the vertical axis differ over region (rows). However, two gray horizontal lines are $5$ years apart in each plot.}
		\label{fig: Topals_Dsplines_e0_boxplot}
	\end{sidewaysfigure}

	Figure \ref{fig: Topals_Dsplines_e0_boxplot} shows boxplots of life expectancy for TOPALS, D-splines, and from the raw data for five exposure sizes, three levels of mortality, and different incorporated demographic knowledge (see supplementary material, Section \ref{sec: sm_additional results}, Figure \ref{fig: sm_Topals_Dsplines_e0_boxplot} for remaining exposure sizes). The figure reveals that for estimating life expectancy TOPALS and D-splines are sensitive to these three dimensions though intensity differs across methods. As exposure size increases, the variability in estimated life expectancy decreases and convergences towards the true value for all combinations of method, region, and incorporated demographic knowledge. At an exposure of $1\,000$ life expectancy varies substantially over the simulation runs. For TOPALS, differences can be more than $40$ years (exposure of $1\,000$ in the high mortality region). The same trend holds true for the life expectancies estimated from the raw data. For the raw data, life expectancy tends to be slightly overestimated in very small populations.
	
	D1 and DLC are more sensitive to the incorporated demographic knowledge than D2 and TOPALS.\footnote{The comparison between TOPALS and D-splines should be made very carefully as the incorporated demographic knowledge is not necessarily comparable.} This effect of sensitivity can be especially observed for very low exposures. For example, the median estimated life expectancy is more than $10$ years apart for dk2 and dk3 in the region with a medium level of mortality for DLC. When the method of estimation is D2 or TOPALS, the median estimated life expectancy is close to the true value. The bias induced by the different incorporated demographic knowledge can also go into different directions. This can again be observed in the region with medium mortality for DLC where life expectancy (median) is overestimated by dk3 but underestimated by dk1 and dk2.
	
	All methods are at least somewhat sensitive to the level of mortality especially for very small populations. However, median estimates of life expectancy are mostly close to the true value for TOPALS and D2 while those D1 and DLC are more variable. Large differences in median estimated life expectancy across different levels of mortality can be observed for very small populations for D1 and DLC. The level of mortality of a region can determine whether life expectancy is over- or underestimated when the incorporated demographic knowledge is fixed. For example, for DLC at an exposure of $1\,000$ with dk1, life expectancy is underestimated in the low mortality region, slightly underestimated in the medium mortality region, and overestimated in the high mortality region. This indicates that the incorporated demographic knowledge \enquote{pulls} the estimates closer together\footnote{Mortality is estimated independently for each region, such that the influence comes solely from the incorporated demographic knowledge.} as little information are available in the data at hand.
	
	\subsection{SVD-model}
	\label{subsec: svd model}
	We implemented automatic model checks to flag potential problems and removed all instances that did not pass these checks. For a combination of exposure size and demographic knowledge, at least $90\%$ of checks should be passed. Otherwise, this combination is excluded from the results section. This is the case for exposures of $75\,000$ and $100\,000$ with dk3 which are not included in the results. 
	Furthermore, we exclude results for exposures of $1\,000\,000$ due to poor performance in the automatic checks. This, however, does not mean that the model does not fit in general. Increasing the number of iterations per chain could help.
	Details on the automatic checks, the number of removed instances as well as further technicalities (number of chains and iterations) are in the supplementary material, Section \ref{sec: sm_diagnostics_SVD}. Results are calculated from the posterior median and $95\%$ credible intervals.

	\subsection*{\normalfont \textit{Ability to estimate log mortality rates}}
	
\begin{sidewaysfigure}
		\centering
		\includegraphics[width = \textwidth]{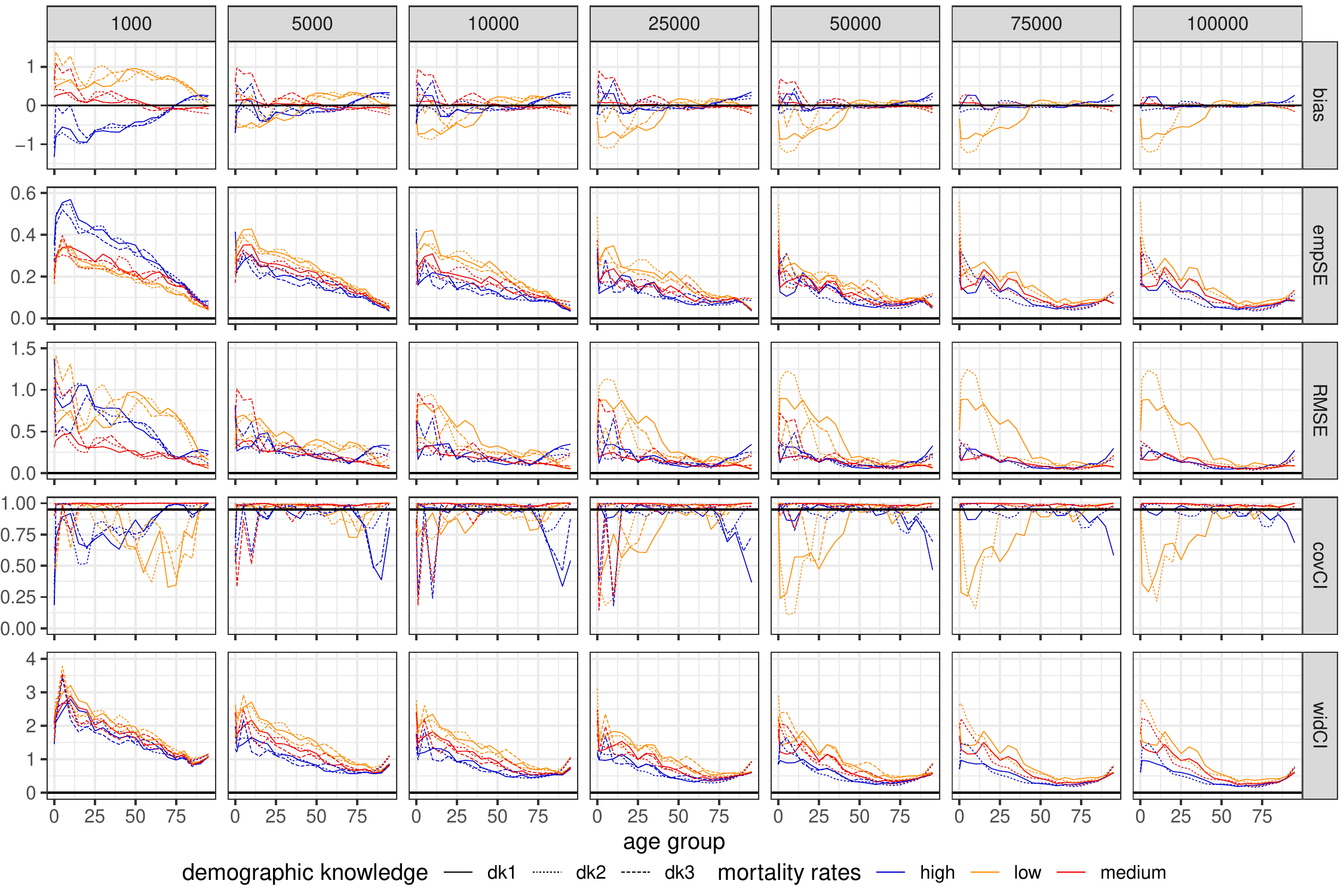}
		\caption{Performance measures for the SVD-model (year $2000$) for three different regions (low ($a = -0.659, b = 0.880$), medium ($a = 0.018$ and $b = 1.035 $), high ($a = 0.693, b = 0.850$) mortality setting), seven exposure sizes, and three types of incorporated demographic knowledge. The idea of this figure is based on \textcite[Figure 1]{MorrisEtAl2019SiM}.}
		\label{fig: SVD_logmx_performance_measures}
	\end{sidewaysfigure}

	Figure \ref{fig: SVD_logmx_performance_measures} reveals that the ability to estimate age-specific log mortality rates varies with age, exposure size, incorporated demographic knowledge and between region. It is evident that bias, empirical standard error, root mean squared error are generally larger and credible intervals wider at younger ages. Coverage of the credible intervals tends to be best around ages of $50$. The figure also allows us to study the influence of exposure size and demographic knowledge across three regions separately.
	
	For different exposure sizes no clear trend is visible except for the width of the confidence intervals which tends to decrease with increasing exposure. There is no clear-cut trend for RMSE and increasing exposures. For example, for the low mortality region (yellow) the RMSE generally increases from an exposure of $1\,000$ to $5\,000$. However, especially for the low mortality region (yellow line) there are pronounced increases in RMSE for exposures larger than $5\,000$. The incorporated demographic knowledge also plays a role within each region (linetype varies with demographic knowledge), especially for the youngest age groups. Lastly, there is no trend across regions with fixed demographic knowledge (different colour but same linetype) which reveals that performance can vary from region to region.
	
	\subsection*{\normalfont \textit{Ability to estimate life expectancy}}
	
\begin{figure}
		\centering
		\includegraphics[width = 0.85\textwidth]{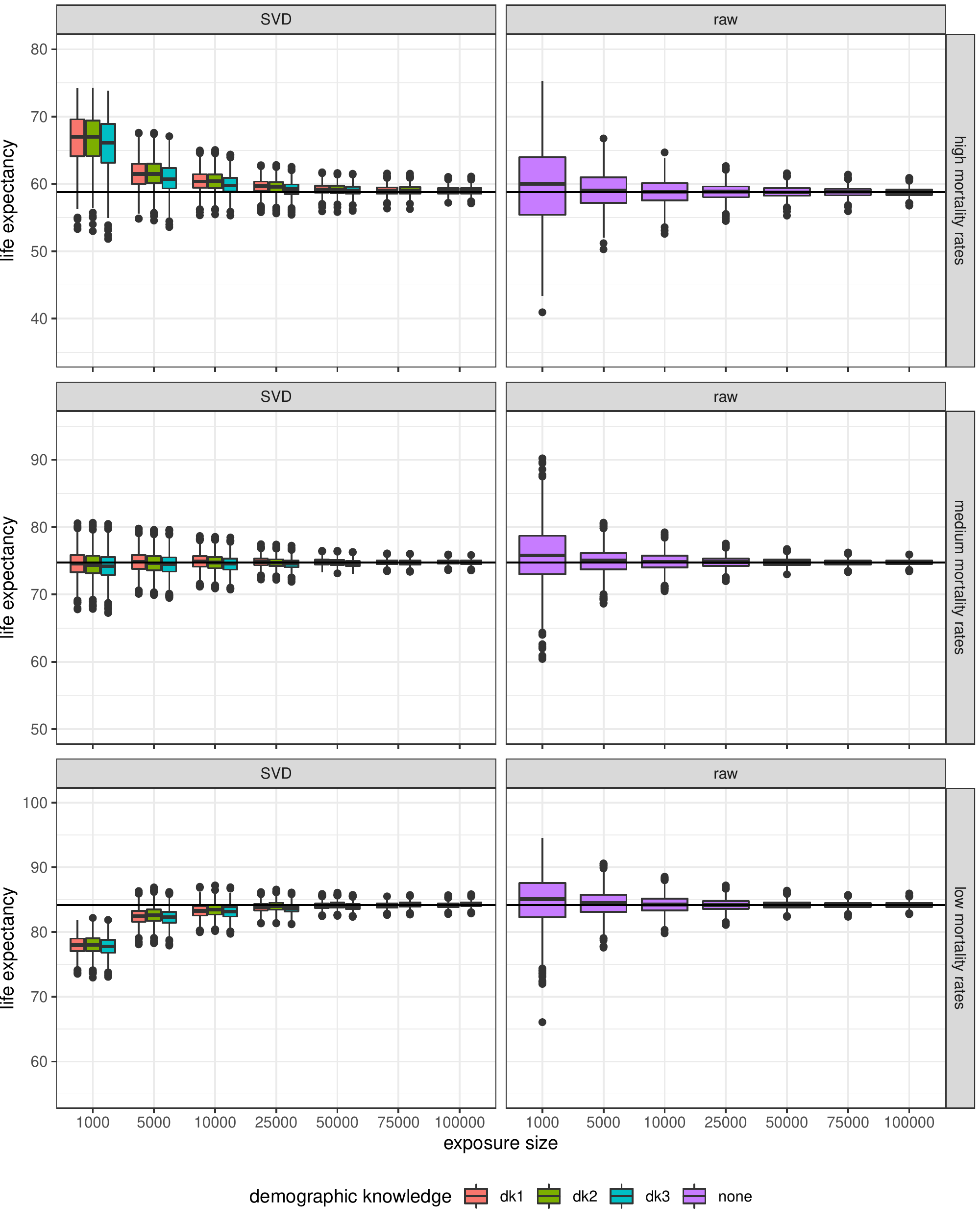}
		\caption{Boxplots of life expectancy in the year $2000$ over all simulation runs for the SVD-model. Estimates from the raw data are shown as well: These are based on all $1\,000$ simulated regions per setting. Boxplots of estimates are based on the number of simulations that passed the automatic checks. The horizontal line is the true life expectancy. Rows correspond to different regions/with different levels of mortality (low ($a = -0.659, b = 0.880$), medium ($a = 0.018$ and $b = 1.035 $), high ($a = 0.693, b = 0.850$) mortality setting). Columns correspond to methods and colours to the incorporated demographic knowledge. Ranges of the vertical axis differ over regions (rows). However, two gray horizontal lines are $5$ years apart in each plot.}
		\label{fig: SVD_e0}
	\end{figure}

	While it is difficult to see clear trends for the log mortality rates, a different picture emerges for life expectancy at birth as shown in Figure \ref{fig: SVD_e0}. For the SVD-model, life expectancy converges towards the true value with increasing exposure. Additionally, variability decreases. While there are slight differences for the different types of demographic knowledge, median estimated life expectancy is generally around the same level. Across regions, there are differences in the quality of results for very small exposures. As such, for the high mortality region life expectancy is overestimated, for the low mortality region it is underestimated and for the medium mortality region around the true value. As a baseline Figure \ref{fig: SVD_e0} also shows results for life expectancy calculated from the raw mortality rates. At small exposures there is a slight overestimation which quickly diminishes with increasing exposure. Variability, however, is larger at small exposures than for the SVD-model.
	
	\section{Discussion \& Outlook}
	\label{sec: discussion}
	
	The paper deals with the evaluation of demographic small-area estimation methods for mortality schedules. We first review demographic small-area estimation methods and then show by means of two simulation studies that there is considerable variability in the performance of methods over different ages. This performance varies with exposure size, incorporated demographic knowledge, and across regions. 
	Before discussing the results of our study and its implications, we would like to acknowledge its three main limitations. 
	The first limitation concerns the design of our simulation study: We simulated death counts for $20$ artificial regions/subpopulations by imposing the age structure of German males and we used the Brass relational model to create the artificial regions/subpopulations. This limits the functional form of the relationship between the male German data and the subpopulations. Additionally, we simulated data for closed age groups only, neglecting any deaths at ages $100$ and above. 
	The second limitation concerns the generalizability of the results. We presented a snapshot of a wealth of simulation results in order to carve out factors that might influence the performance of different methods. This choice comes at the cost of generalizability.
	The third limitation concerns the scope of the chosen methods. \textcite{GonzagaSchmertmann2016RBdEdP} and \textcite{SchmertmannGonzaga2018D} note that researchers from other fields, such as statistical epidemiology have also developed methods for estimating mortality rates and life expectancy in small populations. We, however, deliberately chose the unifying theme of methods that incorporate demographic knowledge about the shape of human mortality schedules.
	
	We find that recent demographic methods differ in their data requirements. As such, frequentist TOPALS regression models and D-splines estimate mortality rates in different regions/subpopulations and over time independently of each other while one Bayesian versions of TOPALS pools across space and the SVD-model across both space and time. This is also reflected in the data requirements and ease of use. Once code is available, (frequentist) TOPALS and D-splines are simple and quick to use. In the simulation study, we reveal that averaging performance measures over several regions or ages may mask underlying variability. As such, we show that all methods are somewhat sensitive to exposure size and the incorporated demographic knowledge across different regions but that there is large variation. While bias tends to be larger and coverage of uncertainty intervals lower for younger ages, there is no clear pattern across methods and regions. In the results presented, D1 and DLC generally exhibit lower variability than D2 and TOPALS which are more accurate on average. Thus, DLC and D1 tend to exhibit more bias which translates into the estimates of life expectancy. For larger exposures this bias for life expectancy tends to diminish. The coverage of uncertainty intervals varies over the ages but tends to be best around the age of $50$. At some ages, especially the youngest, coverage may be zero. Good coverage (i.e.~around 0.95), especially at younger ages tends to come with very wide uncertainty intervals. This may be considered a desired feature: There are many zero death counts at these ages such that we really do not know much.
	
	Based on the presented results, we would advise against using the D2 estimator of D-Splines in (very) small populations as the resulting mortality schedules can be implausible. TOPALS and the SVD-model showed reasonable results for life expectancy while the DLC and D1 estimator tend to be more biased at (very) small populations. Using the raw data to estimate life expectancy seems to be a viable alternative with slight overestimation and huge variability in small populations (see \textcite{EayresWilliams2004JoECH} for more on the overestimation of life expectancy). However, we want to draw attention to two further aspects. First, we want to caution against overinterpreting point estimates. In small populations there is considerable variability in the results and there is likely a limit to conclusions that can be drawn. As \textcite[1383]{SchmertmannGonzaga2018D} write: \enquote{Even with good statistical methods for estimating mortality in very small populations, realistic assessment of uncertainty suggests that drawing meaningful distinctions about the mortality of those populations may be extremely difficult. Demographers should stay humble.}. Second, potential correlations between different regions/subpopulations are ignored when mortality indicators are estimated for each of these aspects separately (e.g.~\textcites{Congdon2009ISR}).
	
	There are several avenues for future research endeavours.
	First, we show that there is considerable variability in the quality of results over different ages which varies with region/subpopulation, incorporated demographic knowledge, and exposure size. However, it would be interesting to study whether more general advice can be formulated.
	Second, from a substantive point of view, it is desirable that male and female mortality schedules do not cross-over at higher ages and that methods produce plausible trends in life expectancy over time. Some demographic methods already try to tackle (parts) of these challenges \parencites{AlexanderEtAl2017D, DharamshiEtAl2021PAA, RauSchmertmann2020DAO}. Future research could evaluate different methods with respect to achieving these goals and contribute new methodologies.
	Lastly, we would like to encourage methodologists to study new methods using a variety of performance measures as well as to avoid averaging over different data generating processes in simulation studies.
	
	\subsubsection*{\small Data Availability}
	\small
	The datasets were derived from sources in the public domain: \url{www.mortality.org}.
	
	\subsubsection*{\small Acknowledgements}
	\small 
	We thank Ugofilippo Basellini, Angela Carollo, Ameer Dharamshi, Ole Hexel, and Rainer Walke for useful comments/technical help. 
	An earlier version of this manuscript was discussed at the roundtable of the Department of Digital and Computational Demography and we thank the participants: Ugofilippo Basellini, Maciej J.~Dańko, Jessica Donzowa, Ana Cristina Gómez Ugarte Valerio, Chia-Jung Tsai, and Xinyi Zhao.
	This work used the Scientific Compute Cluster at GWDG, the joint data center of Max Planck Society for the Advancement of Science (MPG) and University of Göttingen. 
	PG has received funding from the European Research Council (ERC) under the European Union’s Horizon 2020 research and innovation programme (grant agreement No 851485). 
	
	\printbibliography
	
	\newpage
	
	\appendix
	
	\section{True underlying data}
\label{sec: sm_artifical_regions}

This section summarizes the Brass parameters to generate the $20$ artificial regions. It also shows plots of life expectancy and age-specific log mortality rates.

\begin{table}[H]
	\centering
	\caption{Brass parameters applied to data of German males as described in the main body of the paper, resulting in $20$ artificial regions, $i = 1, \dots, 20$.}
	\label{tab: sm_brass_parameters}
	\begin{tabular}{rrr}
		\hline
		$a_i$ 		& $b_i$ 	& region $i$ \\ 
		\hline
		-0.48 		& 0.80 		&   1 \\ 
		0.69 		& 0.85 		&   2 \\ 
		0.04 		& 0.90 		&   3 \\ 
		-0.32 		& 0.82 		&   4 \\ 
		0.20 		& 1.23 		&   5 \\ 
		-0.19 		& 0.72 		&   6 \\ 
		0.33 		& 1.22 		&   7 \\ 
		-0.66 		& 0.88 		&   8 \\ 
		0.02 		& 1.04 		&   9 \\ 
		-0.26 		& 1.05 		&  10 \\ 
		0.62 		& 1.06 		&  11 \\ 
		0.33 		& 1.00 		&  12 \\ 
		0.01 		& 0.85 		&  13 \\ 
		0.37 		& 1.06 		&  14 \\ 
		0.34 		& 0.94 		&  15 \\ 
		-0.67 		& 1.09 		&  16 \\ 
		-0.24 		& 0.77 		&  17 \\ 
		0.24 		& 0.76 		&  18 \\ 
		0.53 		& 1.10 		&  19 \\ 
		-0.06 		& 1.07 		&  20 \\ 
		\hline
	\end{tabular}
\end{table}

\begin{figure}[H]
	\centering
	\includegraphics[width = \textwidth]{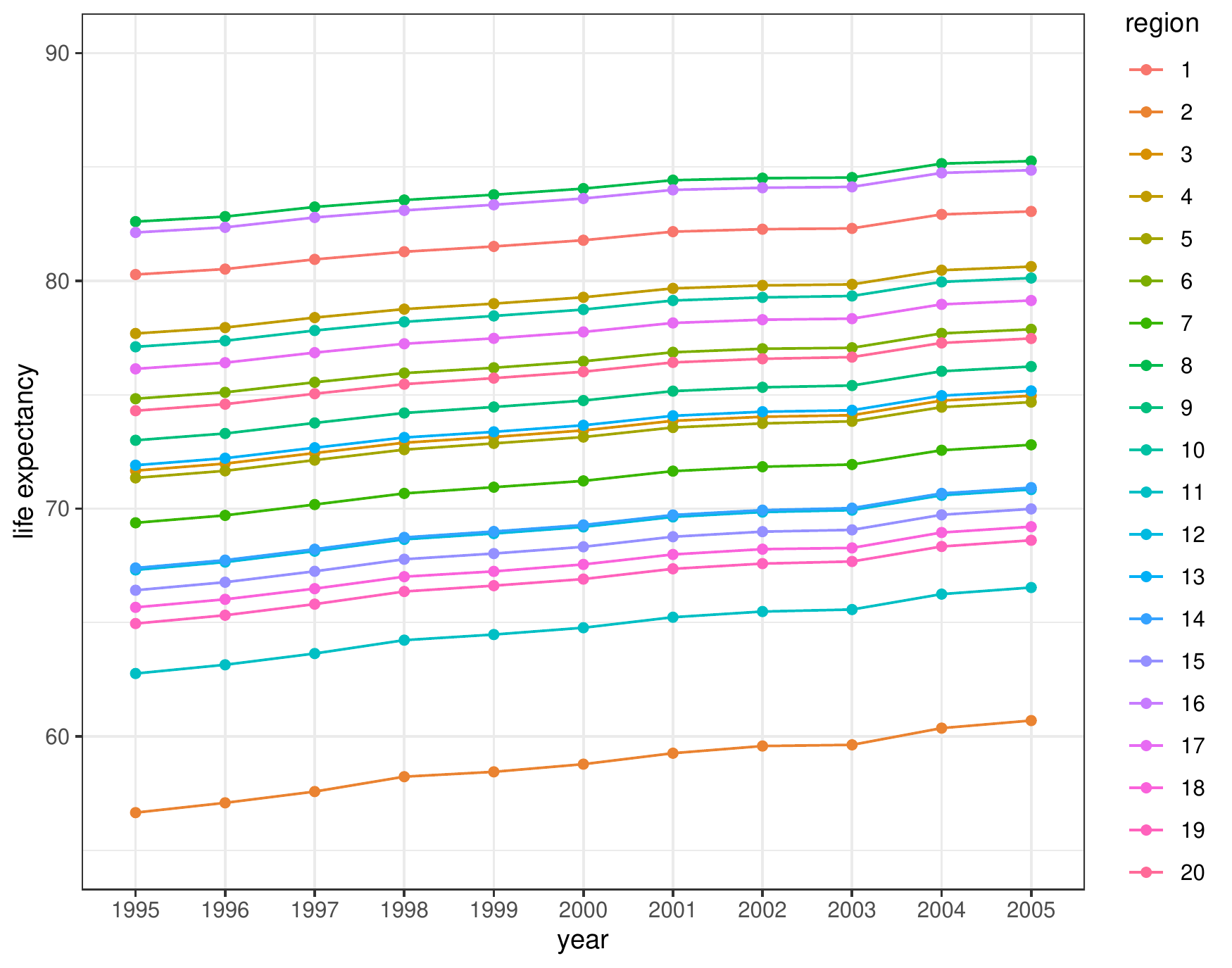}
	\caption{Life expectancy (partial, ages $0-99$) of $20$ artificial regions over time. Regions correspond to the region id's in Table \ref{tab: sm_brass_parameters}.}
	\label{fig: sm_e0_cutoff_true_subpopulations}
\end{figure}

\begin{figure}[H]
	\centering
	\includegraphics[width = \textwidth]{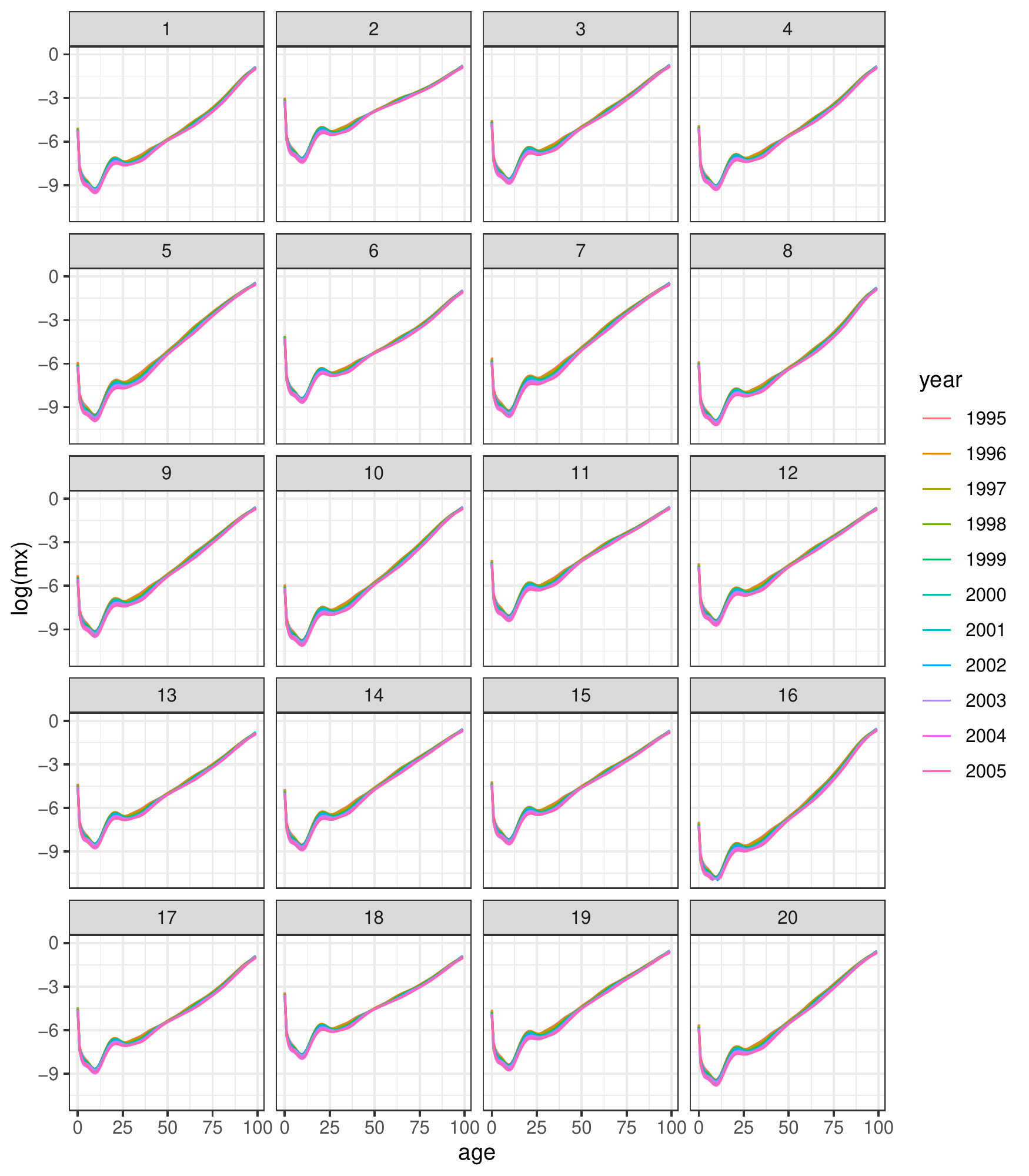}
	\caption{Age-specific log mortality rates over time by region (= $20$ artifical subpopulations). Regions correspond to the region id's in Table \ref{tab: sm_brass_parameters}.}
	\label{fig: sm_logmx_over_time_by_idab_subpopulations}
\end{figure}

\begin{figure}[H]
	\centering
	\includegraphics[width = \textwidth]{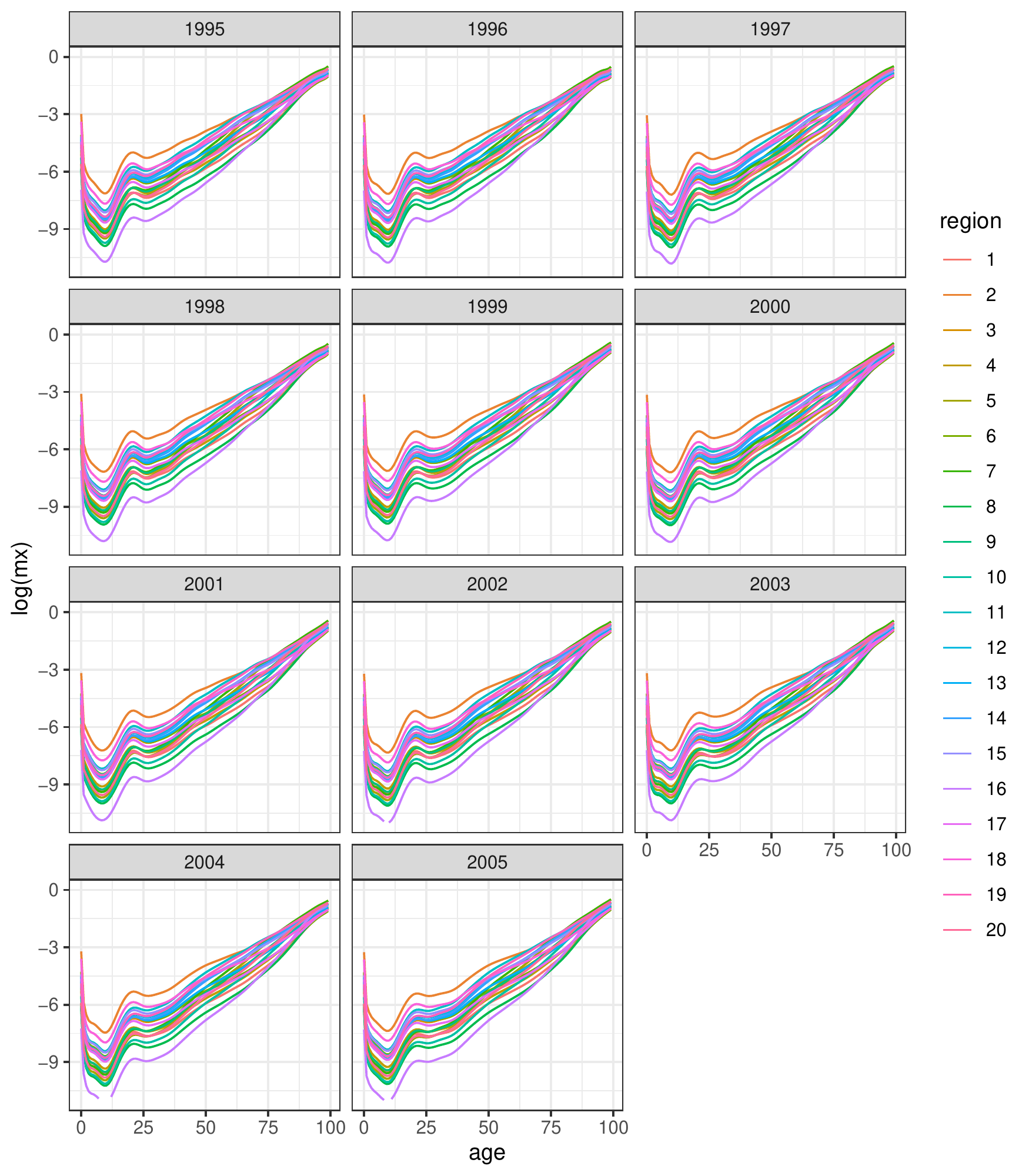}
	\caption{Age-specific log mortality rates of $20$ artificial regions each year. Regions correspond to the region id's in Table \ref{tab: sm_brass_parameters}.}
	\label{fig: sm_logmx_over_time_by_year_subpopulations}
\end{figure}

\newpage

\section{Demographic Knowledge: TOPALS and SVD-model}
\label{sec: dem_know_topals_svd}

\begin{figure}[H]
	\centering
	\includegraphics[width = 0.7\textwidth]{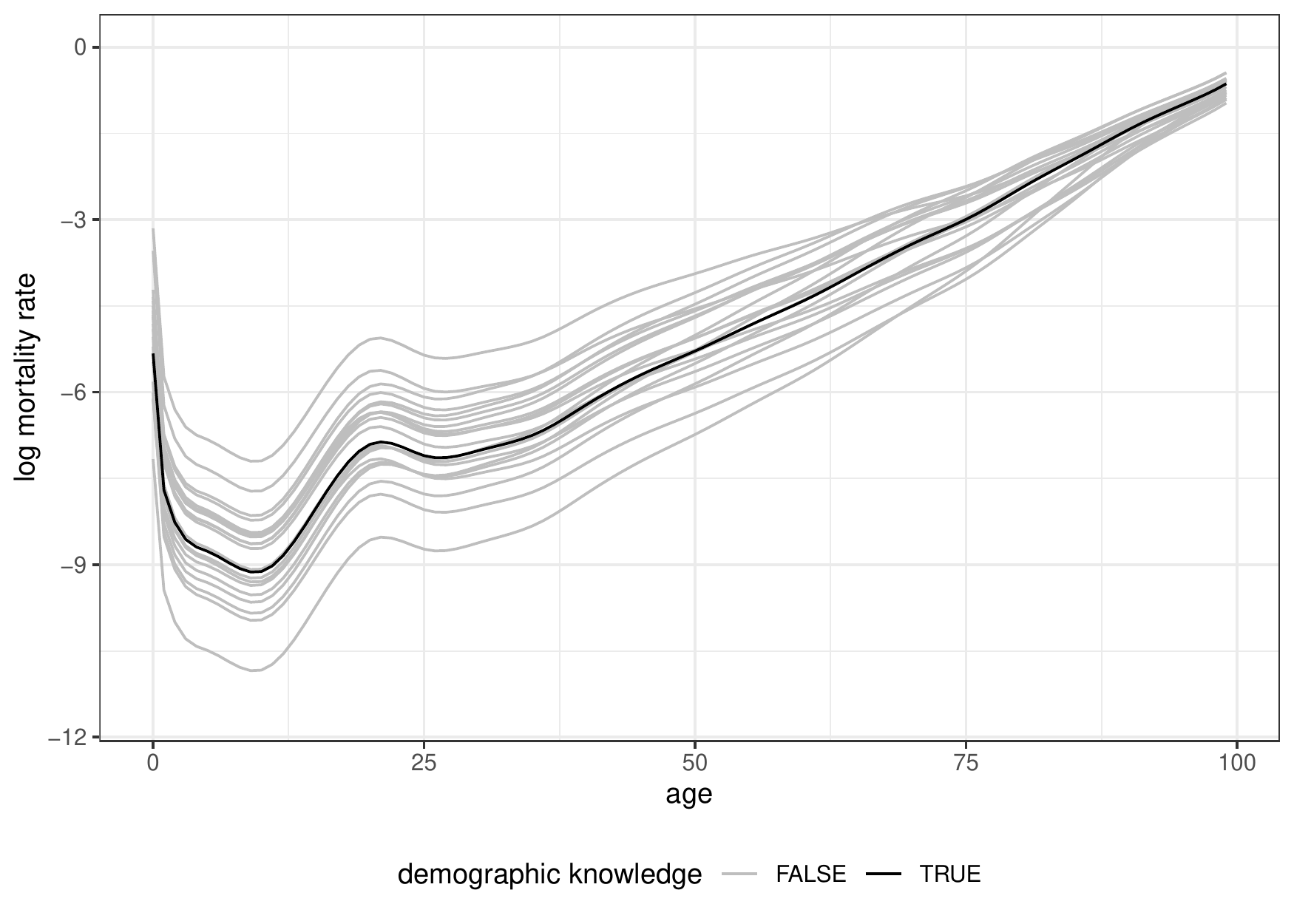}
	\caption{Mortality schedules of demographic knowledge type $1$ (dk1) and the $20$ artificial regions. In black is the schedule used as standard (dk1 here: smoothed male national schedule of Germany for the year $2000$ (HMD code: \enquote{DEUTNP} \parencite{HMD20220711}) and in gray are the $20$ subpopulations/regions (male mortality schedules from which we simulate death counts, year $2000$).}
	\label{fig: sm_dk1_TOPALS_males}
\end{figure}

\begin{figure}[H]
	\centering
	\includegraphics[width = 0.7\textwidth]{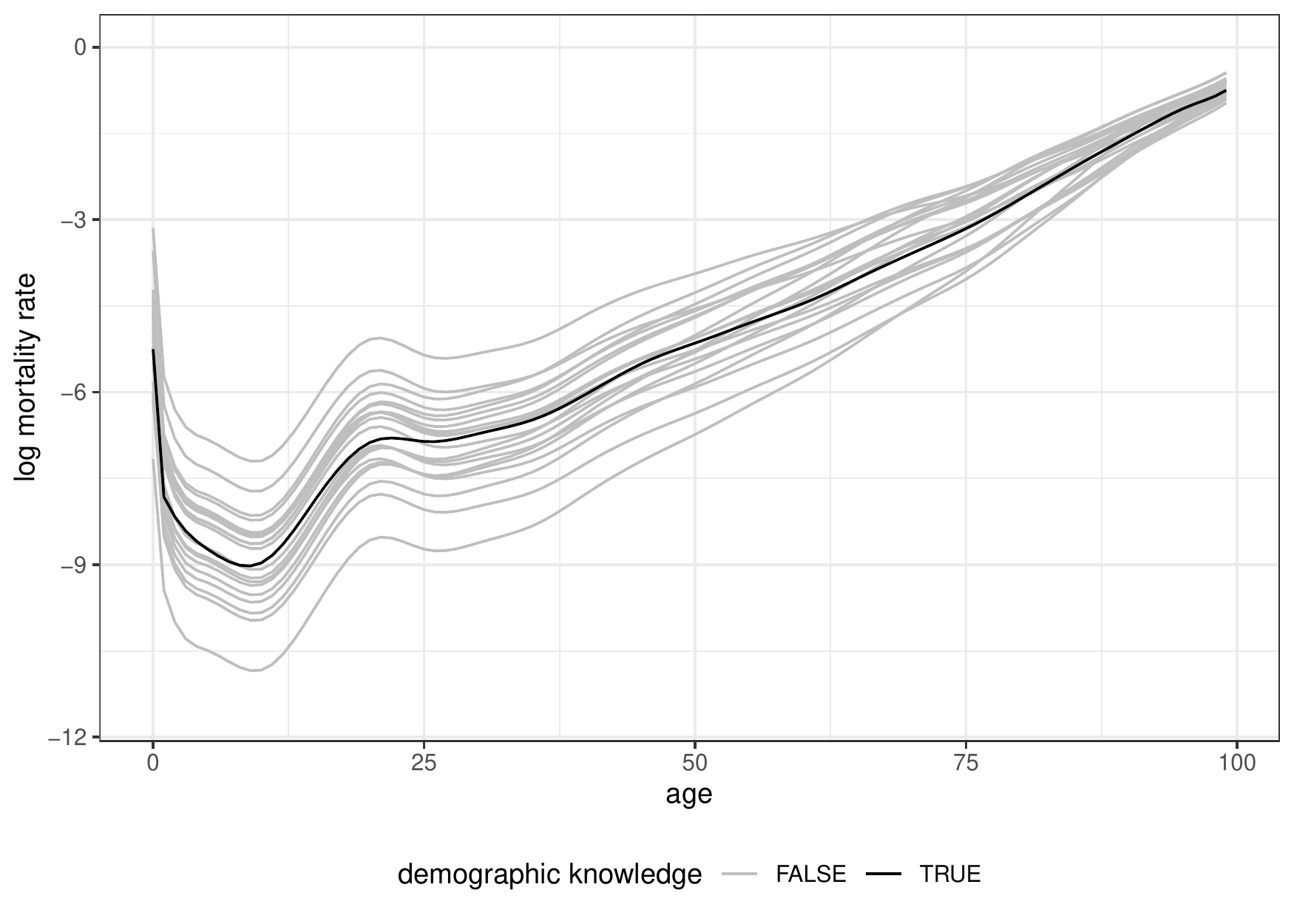}
	\caption{Mortality schedules of demographic knowledge type $2$ (dk2) and the $20$ artificial regions. In black is the schedule used as standard (dk2 here: smoothed male national schedule of France for the year $2000$ (HMD code: \enquote{FRATNP} \parencite{HMD20220711}) and in gray are the $20$ subpopulations/regions (male mortality schedules from which we simulate death counts, year $2000$).}
	\label{fig: sm_dk2_TOPALS_males}
\end{figure}

\begin{figure}[H]
	\centering
	\includegraphics[width = 0.7\textwidth]{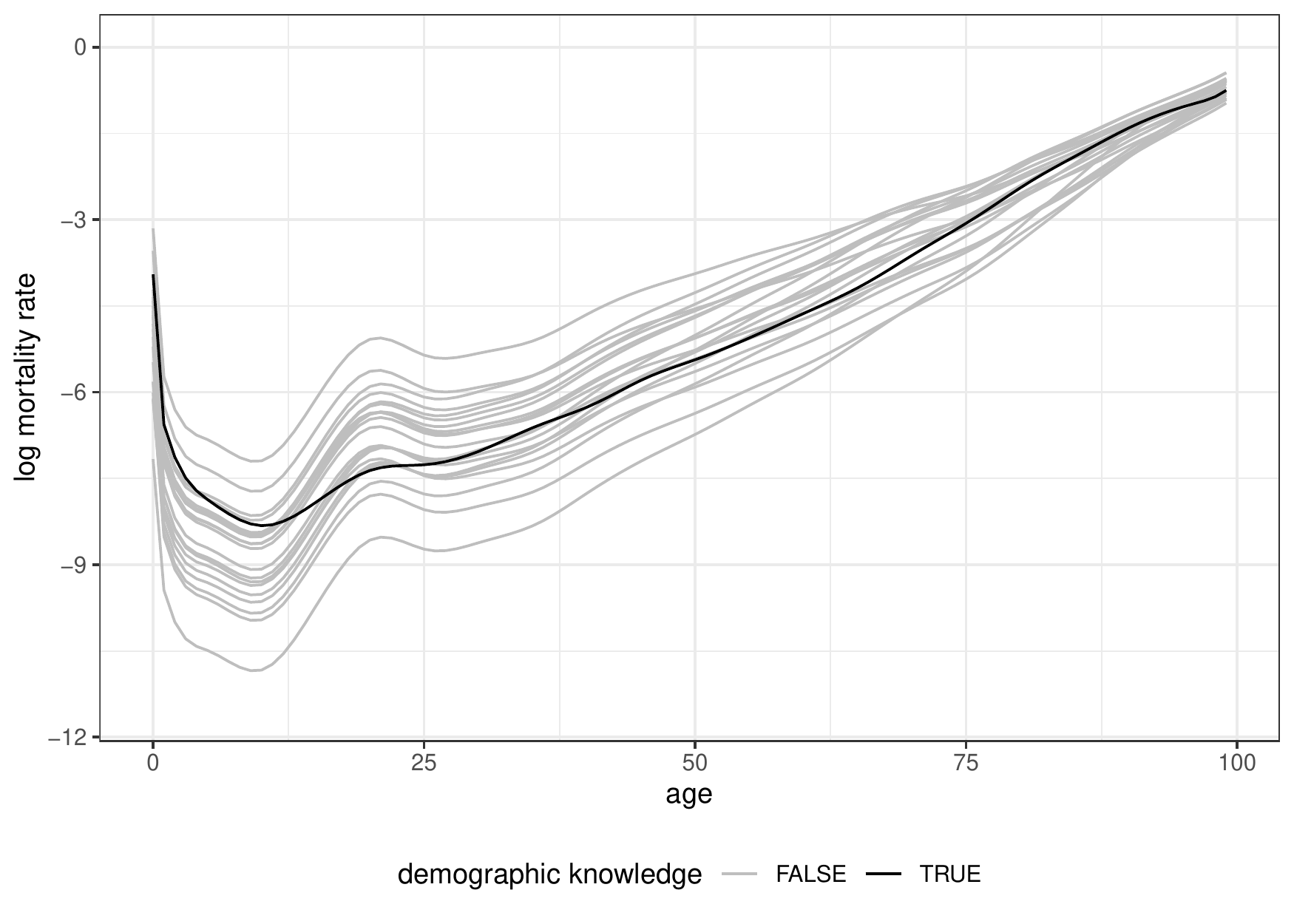}
	\caption{Mortality schedules of demographic knowledge type $3$ (dk3) and the $20$ artificial regions. In black is the schedule used as standard (dk3 here: smoothed female national schedule of France for the year $1965$ (HMD code: \enquote{FRATNP} \parencite{HMD20220711}) and in gray are the $20$ subpopulations/regions (male mortality schedules from which we simulate death counts, year $2000$).}
	\label{fig: sm_dk3_TOPALS_males}
\end{figure}

\begin{figure}[H]
	\centering
	\includegraphics[width = \textwidth]{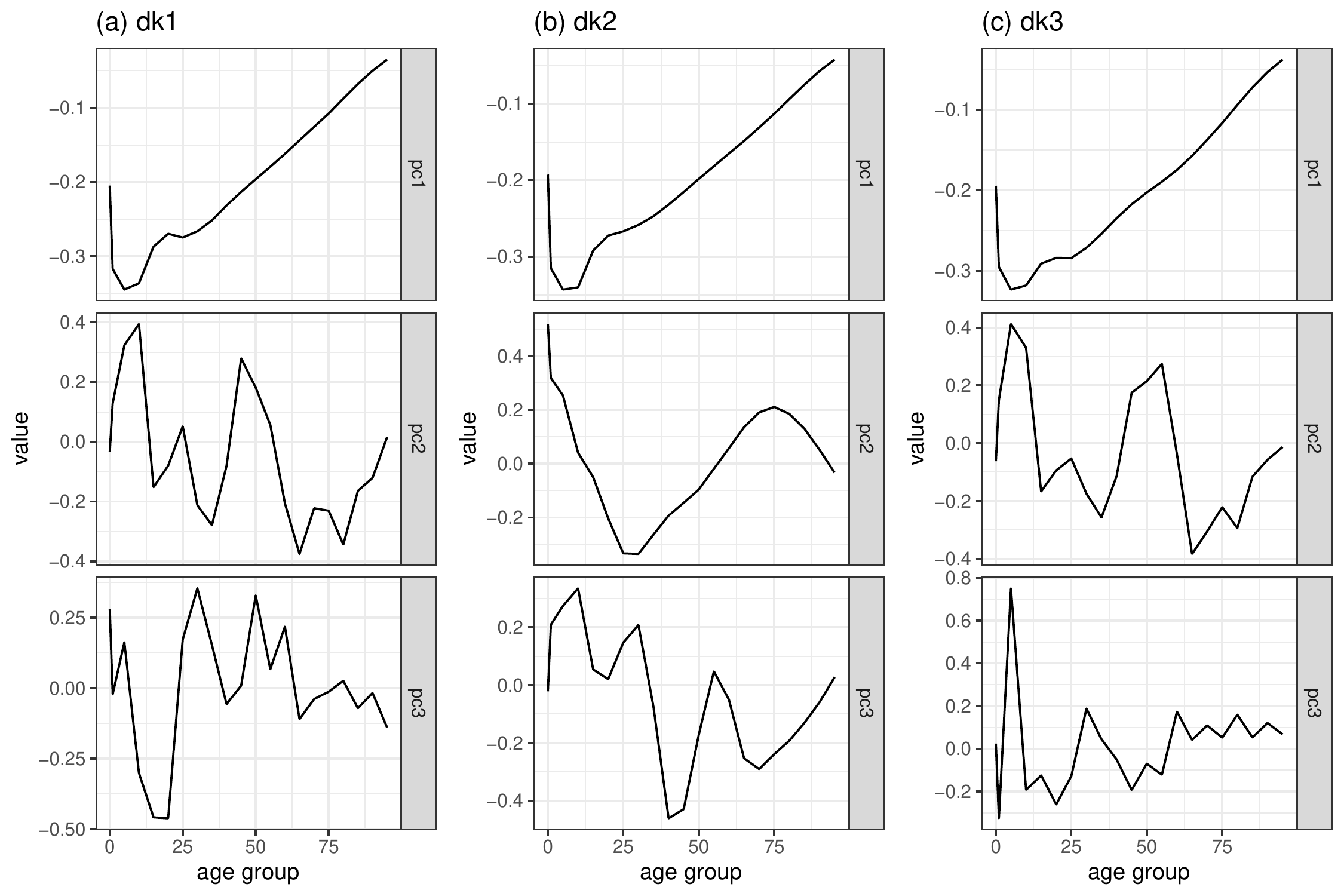}
	\caption{Demographic knowledge for the SVD model, i.e. right singular vectors (\enquote{pc}). dk1 (a) is from the smoothed national time series ($1995 - 2005$) of the reference population (males, Germany, DEUTNP). dk2 (b) is based on all complete mortality schedules from $1970-2018$ excluding Germany (males, HMD codes: \enquote{DEUTNP}, \enquote{DEUTW}, \enquote{DEUTE}). dk3 (c) is constructed as dk1 but from female, instead of male data.}
	\label{fig: sm_plot_pcs_SVD_model}
\end{figure}

\newpage

\section{Performance Measures}
\label{sec: sm_performance_measures}

The bias is defined as the average difference of the estimate $\hat{\theta}$ and the true value $\theta$ \parencite[Table 6]{MorrisEtAl2019SiM}:

\begin{equation}
	\label{eq: bias}
	\widehat{\text{Bias}} = 
	\frac{1}{n_{\text{sim}}} \sum_{j = 1}^{n_{\text{sim}}} \left( \hat{\theta}  - \theta  \right).
\end{equation}

The empirical standard error measures the spread of the estimated values over the simulation runs and is estimated as follows \parencite[Table 6]{MorrisEtAl2019SiM}:

\begin{equation}
	\label{eq: empSE}
	\widehat{\text{EmpSE}} = 
	\sqrt{ \frac{1}{n_{\text{sim}} - 1} \sum_{j = 1}^{n_{\text{sim}}} \left( \hat{\theta}_{j}  - \bar{\theta} \right)^{2} },
\end{equation}

where $ \bar{\theta} = \frac{1}{n_{\text{sim}}} \sum_{j = 1}^{n_{\text{sim}}} \hat{\theta}_{j} $ denotes the average estimate over the simulation runs.

The root mean squared error is defined as \parencite[Table 6]{MorrisEtAl2019SiM}:

\begin{equation}
	\widehat{\text{RMSE}} = 
	\sqrt{ \frac{1}{n_{\text{sim}}} \sum_{j = 1}^{n_{\text{sim}}} \left( \hat{\theta}_{j}  - \theta  \right)^{2} }.
\end{equation}

The coverage of the uncertainty intervals is calculated as \parencite[Table 6]{MorrisEtAl2019SiM}:

\begin{equation}
	\widehat{\text{CovCI}} = 
	\frac{1}{n_\text{sim}} \sum_{j = 1}^{n_{\text{sim}}} \mathbbm{1} \left( \hat{\theta}_{j}^{\text{ low}} \leq \theta \leq \hat{\theta}_{j}^{\text{ upp}} \right),
\end{equation}

where $\mathbbm{1}\left(\cdot\right)$ denotes the indicator function, $ \hat{\theta}_{j}^{\text{ low}} $ and $ \hat{\theta}_{j}^{\text{ upp}}$ are the lower and upper bound of the confidence interval, respectively. Lastly, we calculate the width of the confidence intervals as
\begin{equation}
	\widehat{\text{WidCI}} = 
	\frac{1}{n_\text{sim}} \left( \vert \log \left( \hat{\theta}_{j}^{\text{ upp}} \right)  - \log \hat{\theta}_{j}^{\text{ low}} \vert \right).
\end{equation}

We calculate the performance measures on the log mortality rates, i.e.~$\hat{\theta}_{j} = \log \left( \hat{m}_x^{it,j} \right)$ and $ \theta =  \log \left( m_x^{it} \right) $.

\newpage

\section{SVD-model}
\label{sec: sm_svd_model_formulation}

Let $D_{g,a,t}$ be the number of deaths and $N_{g,a,t}$ the exposure in age group $g$, region $a$, and year $t$. The corresponding mortality rate is denoted as $m_{g,a,t}$. We use three principal components, $p = 1, 2, 3$. 

\begin{equation}
	D_{g,a,t} \sim \mathcal{N} \left( m_{g,a,t} N_{g,a,t} \right)
\end{equation}

\begin{equation}
	\log \left( m_{g,a,t} \right) 
	= \beta_{1,a,t} V_{1,g} + \beta_{2,a,t} V_{2,g} + \beta_{3,a,t} V_{3,g} + u_{g,a,t}
\end{equation}

\begin{equation}
	\beta_{p,a,t} \sim \mathcal{N} \left( \mu_{\beta_{p,t}}, \sigma_{\beta_{p,t}} \right)
\end{equation}

\begin{equation}
	\mu_{\beta_{p,t}} \sim \mathcal{N} \left( 2 \cdot \mu_{\beta_{p,t-1}} - \mu_{\beta_{p,t-2}}, \sigma_{\beta_{\mu_{p}}} \right)
\end{equation}

\begin{equation}
	\sigma_{\beta_{\mu_{p}}} \sim \mathcal{LN} \left( -1.5, 0.5 \right)
\end{equation}

\begin{equation}
	\sigma_{\beta_{p,t}} \sim \mathcal{N}_{+} \left( 0, 1 \right)
\end{equation}

\begin{equation}
	u_{g,a,t} \sim \mathcal{N} \left( 0, \sigma_{g} \right)
\end{equation}

\begin{equation}
	\sigma_{g} \sim \mathcal{N}_{+} \left( 0, 0.25 \right)
\end{equation}

\newpage

\section{Additional Results}
\label{sec: sm_additional results}

\begin{sidewaysfigure}
	\centering
	\includegraphics[width = \textwidth]{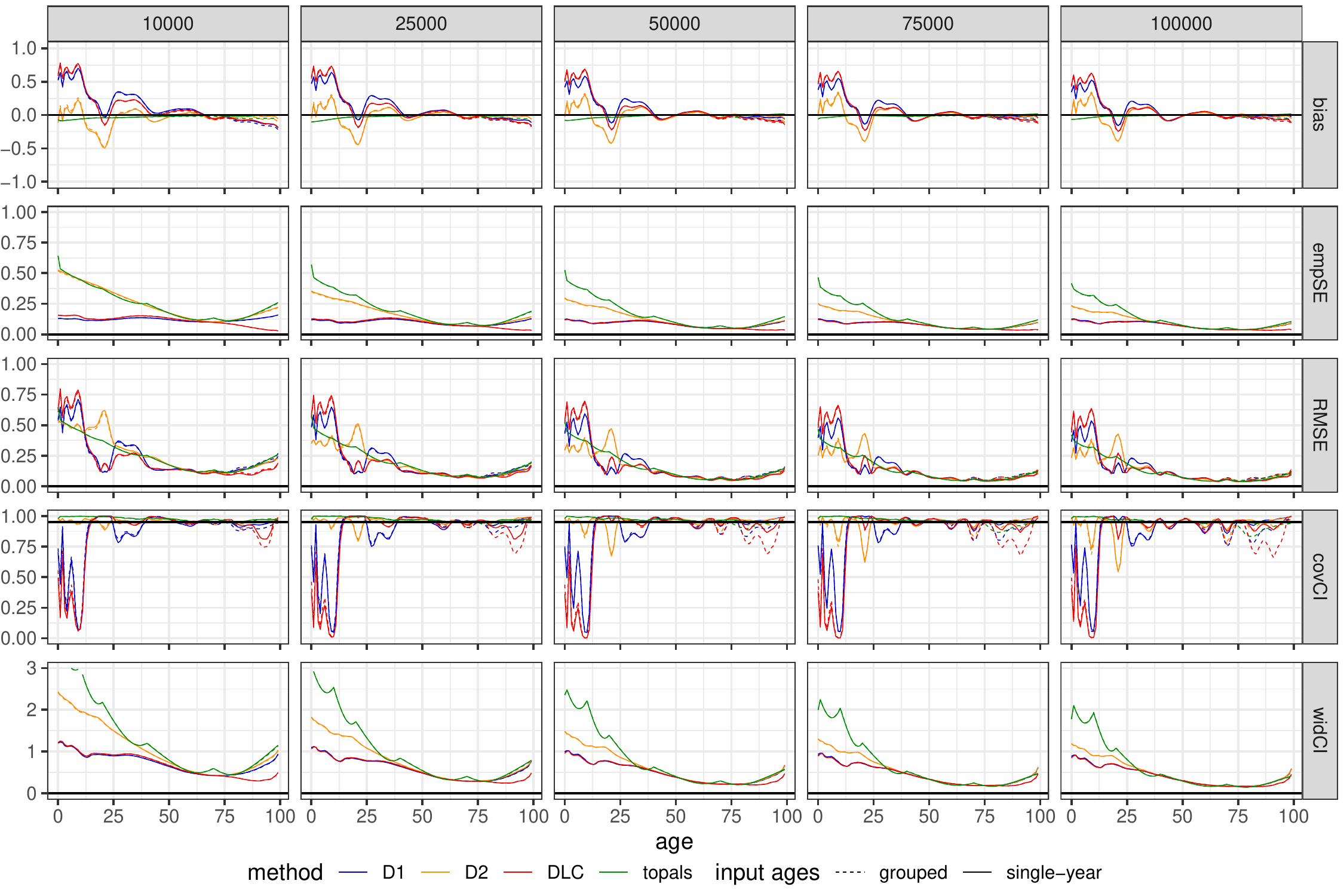}
	\caption{Performance measures for TOPALS and D-splines (D1, D2, DLC) over five different exposures sizes ($50\,000$ and $100\,000$ overlap with Figure \ref{fig: Topals_Dsplines_over_exposures_idab9_male} in the main paper as reference). Results shown for age-grouped (dashed line) and single year age (solid line) input data. Results shown are for the year $2000$ for one fixed region (similar to schedule of German males in the year $2\,000$ with Brass parameters $a = 0.018$ and $b = 1.035 $) and the canonical demographic knowledge (TOPALS: German males for the year $2000$. D-splines: All complete male mortality schedules in the HMD since $1970$.). The empirical standard error is not comparable across methods in the presence of bias \parencite{MorrisEtAl2019SiM}. Additionally, some lines are cut-off on the vertical axis due to very large values (e.g.~width of confidence intervals for D2 at small exposures). The idea of this figure is based on \textcite[Figure 1]{MorrisEtAl2019SiM}.}
	\label{fig: sm_Topals_Dsplines_over_exposures_idab9_male}
\end{sidewaysfigure}

\begin{sidewaysfigure}
	\centering
	\includegraphics[width = \textwidth]{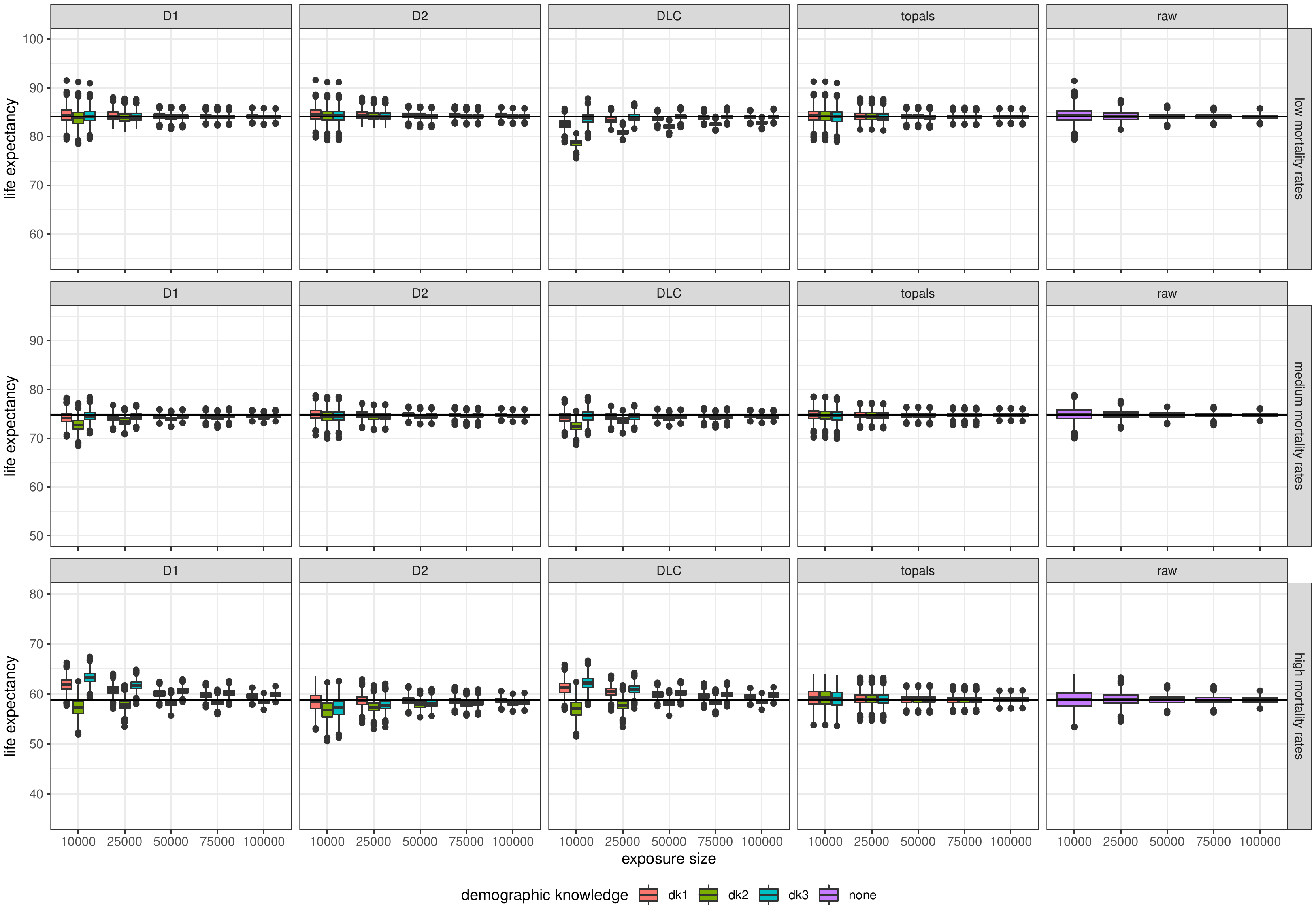}
	\caption{Boxplots of partial life expectancy (ages $0-99$) over all simulation runs for TOPALS, D-splines (D1, D2, DLC) for single year age input data for the year $2000$. Estimates from the raw data are shown as well. Exposure sizes complement those in the main paper in Figure \ref{fig: Topals_Dsplines_e0_boxplot} ($50\,000$ and $100\,000$ overlap with Figure 1 in the main paper as reference). The horizontal line is the true life expectancy. Rows correspond to different regions/subpopulations with different levels of mortality (low ($a = -0.659, b = 0.880$), medium ($a = 0.018$ and $b = 1.035 $), high ($a = 0.693, b = 0.850$) mortality setting). Columns correspond to methods and colors to the incorporated demographic knowledge. Ranges of the vertical axis differ over region (rows). However, two gray horizontal lines are $5$ years apart in each plot.}
	\label{fig: sm_Topals_Dsplines_e0_boxplot}
\end{sidewaysfigure}

\newpage

\section{SVD-model: technical details \& diagnostics}
\label{sec: sm_diagnostics_SVD}

We ran each model with four chains and the number of iterations depending on the exposure size. For exposures of $1\,000$ and $5\,000$, there were a total of $3\,000$ iterations including $500$ warm-up iterations per chain. For larger exposures ($10\,000$ and more), there were $6\,000$ iterations including $1\,000$ warm-up iterations per chain.

\begin{sidewaystable}
	\centering
	\caption{Diagnostics for SVD-model. Numbers indicates the number of times automatic checks were passed. We excluded results when not all automatic checks were passed. Thus, column \textit{checks passed} is the number of simulation runs included in the analysis. Column \textit{completed runs} refers to the number of completed runs: For dk1 and an exposure of $100\,000$ there was one run that did not finish within the time limit of 12 hours. Furthermore, we did not complete all simulations for exposures of $1\,000\,000$ due to high rates of failed automatic checks. This does not mean that the model does not fit in general. Increasing the number of iterations per chain might help but is beyond the scope of this paper.}
	\label{tab: sm_diagnostics_SVD}
	\begin{tabular}{rlrrrrrrrr}
		\hline
		exposure & dk & checks passed & Rhat & bulk ESS & tail ESS & divergences & treedepth & BFMI & completed runs \\ 
		\hline
		1000 & dk1 & 979 & 998 & 1000 & 987 & 992 & 1000 & 1000 & 1000 \\ 
		1000 & dk2 & 974 & 997 & 999 & 984 & 990 & 1000 & 1000 & 1000 \\ 
		1000 & dk3 & 986 & 1000 & 1000 & 994 & 992 & 1000 & 1000 & 1000 \\ 
		5000 & dk1 & 976 & 997 & 1000 & 981 & 997 & 1000 & 1000 & 1000 \\ 
		5000 & dk2 & 980 & 995 & 1000 & 992 & 992 & 1000 & 1000 & 1000 \\ 
		5000 & dk3 & 967 & 990 & 999 & 977 & 997 & 1000 & 1000 & 1000 \\ 
		10000 & dk1 & 981 & 1000 & 1000 & 1000 & 981 & 1000 & 1000 & 1000 \\ 
		10000 & dk2 & 997 & 1000 & 1000 & 1000 & 997 & 1000 & 1000 & 1000 \\ 
		10000 & dk3 & 998 & 1000 & 1000 & 1000 & 998 & 1000 & 1000 & 1000 \\ 
		25000 & dk1 & 995 & 998 & 1000 & 999 & 998 & 1000 & 1000 & 1000 \\ 
		25000 & dk2 & 996 & 998 & 1000 & 1000 & 998 & 1000 & 1000 & 1000 \\ 
		25000 & dk3 & 992 & 997 & 1000 & 998 & 997 & 1000 & 1000 & 1000 \\ 
		50000 & dk1 & 980 & 984 & 1000 & 999 & 997 & 1000 & 1000 & 1000 \\ 
		50000 & dk2 & 990 & 991 & 1000 & 999 & 1000 & 1000 & 1000 & 1000 \\ 
		50000 & dk3 & 987 & 988 & 999 & 1000 & 999 & 1000 & 1000 & 1000 \\ 
		75000 & dk1 & 981 & 982 & 1000 & 1000 & 999 & 1000 & 1000 & 1000 \\ 
		75000 & dk2 & 985 & 986 & 1000 & 999 & 1000 & 1000 & 1000 & 1000 \\ 
		75000 & dk3 & 631 & 731 & 724 & 913 & 999 & 1000 & 1000 & 1000 \\ 
		100000 & dk1 & 933 & 933 & 999 & 999 & 999 & 999 & 999 & 999 \\ 
		100000 & dk2 & 908 & 910 & 1000 & 998 & 1000 & 1000 & 1000 & 1000 \\ 
		100000 & dk3 & 490 & 575 & 606 & 657 & 1000 & 970 & 1000 & 1000 \\ 
		1000000 & dk1 &  44 &  52 &  64 &  65 & 842 & 766 & 842 & 842 \\ 
		1000000 & dk2 & 249 & 328 & 524 & 564 & 852 & 748 & 852 & 852 \\ 
		1000000 & dk3 &   2 &   6 &  13 &  19 & 634 & 531 & 634 & 634 \\ 
		\hline	
		\multicolumn{10}{l}{
			\begin{tabularx}{0.95\textwidth}{X}
				\textit{Note:} Rhat, bulkESS and tailESS calculated using function \texttt{rstan::monitor()} \parencite{rstan.2.21.8}. Following recommendations in \textcite{VehtariEtAl2021BA}, we require $\widehat{R} < 1.01, \text{bulkESS} > 400$, and $\text{tailESS} > 400$ for each parameter. Number of divergent transitions must be $0$ (\texttt{rstan::get\_num\_divergent()}), maximum treedepth(=12) should not be reached (\texttt{rstan::get\_num\_max\_treedepth()}), and there should be no low BFMI chains (\texttt{rstan::get\_low\_bfmi\_chains()}).
			\end{tabularx}
		} 
	\end{tabular}
\end{sidewaystable}

\end{document}